\documentclass[a4paper,10pt]{article}

\usepackage{amsmath,amssymb}
\usepackage[usenames]{color}
\usepackage{colortbl}
\usepackage{graphicx}
\usepackage{xspace}
\usepackage{multirow}
\newcommand{\ssqtt}{\ensuremath{\sin^2(2\theta_{13})}\xspace}
\newcommand{\Nbins}{\ensuremath{N_{\text{b}}}\xspace}
\newcommand{\Ndet}{\ensuremath{N_{\text{d}}}\xspace}
\newcommand{\Nreac}{\ensuremath{N_{\text{r}}}\xspace}

\newcommand{\Ns}{\ensuremath{N_{\text{s}}}\xspace}
\newcommand{\sabs}{\ensuremath{\sigma_{\text{abs}}}\xspace}
\newcommand{\srel}{\ensuremath{\sigma_{\text{rel}}}\xspace}
\newcommand{\sscl}{\ensuremath{\sigma_{\text{scl}}}\xspace}
\newcommand{\sscla}{\ensuremath{\sigma_{\text{scl}}^{\text{abs}}}\xspace}
\newcommand{\ssclr}{\ensuremath{\sigma_{\text{scl}}^{\text{rel}}}\xspace}
\newcommand{\sshp}{\ensuremath{\sigma_{\text{shp}}}\xspace}
\newcommand{\spwr}{\ensuremath{\sigma_{\text{pwr}}}\xspace}

\newcommand{\sptp}{\ensuremath{\sigma_{\text{pair}}}\xspace}
\newcommand{\sbkg}{\ensuremath{\sigma_{B_n^D}}\xspace}
\newcommand{\Nfis}{\ensuremath{N_{\ell}^{\text{fis}}}\xspace}
\newcommand{\Efis}{\ensuremath{E_{\ell}^{\text{fis}}}\xspace}

\newcommand{\scmp}{\ensuremath{\sigma_{\text{cmp}}}\xspace}
\newcommand{\nueb}{\ensuremath{\overline{\nu}_e}\xspace}
\newcommand{\dd}{\ensuremath{\text{d}}\xspace}
\newcommand{\Enu}{\ensuremath{E_{\nu}}\xspace}
\newcommand{\Epos}{\ensuremath{E_{e^+}}\xspace}
\newcommand{\Evis}{\ensuremath{E_{\text{vis}}}\xspace}
\newcommand{\Uf}{\ensuremath{{}^{235}\text{U}}\xspace}
\newcommand{\Uh}{\ensuremath{{}^{238}\text{U}}\xspace}
\newcommand{\Pn}{\ensuremath{{}^{239}\text{Pu}}\xspace}
\newcommand{\Po}{\ensuremath{{}^{241}\text{Pu}}\xspace}

\newcommand{\eg}{\emph{e.g.}}
\newcommand{\etc}{\emph{etc.}\xspace}
\newcommand{\mwe}{m.w.e.\xspace}

\newcommand{\red}{\color{red}}

\newcommand{\org}{\color{Orange}}
\newcommand{\hlc}[1]{#1}
\definecolor{gray}{rgb}{.4,.4,.4}

\begin{document}

\title{A unified analysis of the reactor neutrino program\\
       towards the measurement of the $\theta_{13}$ mixing angle}
\author{\textsc{G. Mention}$\,^{(\text{a})}$, \textsc{Th. Lasserre}$\,^{(\text{a,b})}$, \textsc{D. Motta}$\,^{(\text{a})}$\\
$^{(\text{a})}$\,DAPNIA/SPP, CEA Saclay, 91191 Gif sur Yvette, France\\
$^{(\text{b})}$\,Laboratoire Astroparticule et Cosmologie (APC), Paris, France}
\date{\today}
\maketitle

\begin{abstract}
We present in this article a detailed quantitative discussion of the measurement of the leptonic mixing angle $\theta_{13}$ through currently scheduled reactor neutrino oscillation experiments. We thus focus on Double~Chooz (Phase~I \& II), Daya~Bay (Phase~I \& II) and RENO experiments. We perform a unified analysis, including systematics, backgrounds and accurate experimental setup in each case. Each identified systematic error and background impact has been assessed on experimental setups following published data when available and extrapolating from Double~Chooz acquired knowledge otherwise.
After reviewing the experiments, we present a new analysis of their sensitivities to \ssqtt and study the impact of the different systematics based on the pulls approach. Through this generic statistical analysis we discuss the advantages and drawbacks of each experimental setup.
\end{abstract}

\section{Experimental context}
\label{sec:introduction}

Over the last years the phenomenon of neutrino flavor conversion induced by nonzero neutrino masses has been demonstrated by experiments with solar~\cite{solar,SKsolar,sno}, atmospheric~\cite{SKatm}, reactor~\cite{kamland1,kamland} and accelerator neutrinos~\cite{K2K,MINOS}. Neutrino oscillation, that can be described by the Pontecorvo--Maki--Nakagawa--Sakata (PMNS) mixing matrix~\cite{BPont57}, is the current best mechanism to explain the data. Considering only the three known families, the neutrino mixing matrix is parameterized by the three mixing angles ($\theta_{12}$, $\theta_{23}$, $\theta_{13}$) and a possible $\delta$ CP violation phase. The angle $\theta_{12}$ has been measured to be large ($\sin^{2}(2\theta_{12})\simeq 0.8$), the angle $\theta_{23}$ has been measured to be close to maximum ($\sin^{2}(2\theta_{23})\gtrsim 0.9$); but the last angle $\theta_{13}$ has only been upper bounded $\ssqtt\lesssim 0.15$ at~90~\%~C.L., by the CHOOZ reactor experiment~\cite{chooz,limits}. These achievements have now shifted the field of neutrino oscillation physics into a new era of precision measurements. Next generation experiments are underway all around the world to further pin down the values of the oscillation parameters of both solar and atmosheric driven oscillations. Currently the most important task, for the experimentalists, is the determination of the last oscillation through the measurement of the unknown mixing angle $\theta_{13}$. An improved sensitivity on $\theta_{13}$ is not only important for the understanding of neutrino oscillations, but also to open up the possibility of observing CP-violation in the lepton sector if the $\theta_{13}$~driven oscillation is discovered by the forthcoming experiments.

In order to improve the CHOOZ constraint on $\theta_{13}$ at least two identical unsegmented liquid scintillator neutrino detectors close to a nuclear power plant (NPP) are required. The near detector(s) located a few hundred meters away from the reactor cores monitor the unoscillated \nueb~flux. The far detector(s) is(are) located at a distance between 1~and 2~km, searching for a departure from the $1/L^2$ behavior induced by oscillations. Experimental errors are being partially cancelled when using identical detectors. The goal is to achieve an overall effective systematic error of less than $1~\%$~\cite{RWhitePaper,Mention}.

Three experiments have received partial or full approval to perform such a measurement in a near future: Double~Chooz in France~\cite{DChooz}, Daya~Bay in China~\cite{DBProposal} and RENO in Korea~\cite{RENO}.
In addition, a project is under study at the Angra power plant in Brazil to further improve the sensitivity of the measurement on a longer time scale~\cite{Angra}. Furthermore the Japanese KASKA collaboration is promoting the reactor neutrino oscillation physics~\cite{KASKA}.

\section{Neutrino oscillation at reactor and $\mathbf{\theta_{13}}$}
\label{sec:theta13reacor}

Fission reactors are prodigious sources of electron antineutrinos which have a continuous energy spectrum up to about 10~MeV. For $E_{\nueb} > E_{\text{thr}} =  1.806~\text{MeV}$ they can be detected though the $\nueb+p\rightarrow e^+ +n$ reaction using the delayed coincidence technique, where an electron antineutrino interacts with a free proton in a tank containing a target volume filled with Gd loaded liquid scintillator. The positron and the resultant annihilation gamma-rays are detected as a prompt signal while the neutron slows down and then thermalizes in the liquid scintillator before being captured by a hydrogen or gadolinium nucleus. The excited nucleus then emits gamma rays which are detected as the delayed signal. Electron antineutrinos energy is derived from the measured visible energy from positron energy loss and annihilation,
\begin{equation}
\label{eq:Evis}
\Evis=\Epos+m_e\simeq\Enu-\Enu^{\text{thr}}+2\,m_e\;.
\end{equation}

The \nueb spectrum is calculated from  measurements of the beta decay spectra of \Uf, \Pn and \Po~\cite{nuspectrum} after fissioning by thermal neutrons and theoretical \Uh~spectrum, since no data are available for~\Uh \footnote{\Uh fissions only with fast neutrons. Theoretical predictions are computed by summing all known beta decay processes contributing.}.
As a nuclear reactor operates, the fission element proportions evolve in time (the so-called burn-up). Since we are interested here on long term interpretation of the data for oscillation search, we will use an average fuel composition for a reactor cycle corresponding to
\begin{equation}
  \label{eq:mean_composition}
  \Uf\;(55.6~\%),\;\Pn\;(32.6~\%),
  \;\Uh\;(7.1~\%)\;\text{and}\;\Po\;(4.7~\%)\;.
\end{equation}
The mean energy release per fission, $\left<E_f\right>$, is then 205~MeV and the energy weighted cross section for $\nueb p \rightarrow e^+ n$ amounts to $\left<\sigma_f\right>= 6\;10^{-43}$~cm$^2$ per fission. Let us introduce a new luminosity unit, called the r.n.u. (for reactor neutrino unit) and defined as $1~\text{r.n.u.}=0.197\;10^{60}~\text{MeV}$. With this unit, an experiment taking data for $T$~years with a total NPP (nuclear power plant) thermal power of $P$~GW and with $N\;10^{30}$~free protons inside the target has a luminosity ${\cal L}=T\,P\,N~\text{r.n.u.}$. The event number, $N(L)$, at a distance $L$ from the source, assuming no\,-\,oscillation, can be quickly assessed with
\begin{equation}
\label{eq:rnu}
 N(L)=\frac{\left<\sigma_f\right>}{4\pi\left<E_f\right>}\frac{\cal L}{L^2}\simeq 4.6\;10^{9}\;
      \left(\frac{T}{1~\text{yr}}\right)
      \left(\frac{P}{1~\text{GW$_{\text{th}}$}}\right)
      \left(\frac{N}{10^{30}}\right)
      \left(\frac{1~\text{m}}{L}\right)^2
\end{equation}
For the full antineutrino reactor energy spectra simulation, we follow Vogel's analytical parameterization~\cite{Vogel}, based on second order \Enu polynomials. Higher order parameterizations~\cite{HuberSchwetz} give very comparable results and do not require a specific attention for our aim in this article. The antineutrino event rates per energy range is then computed according to the mean reactor core composition~(\ref{eq:mean_composition}) and experimental site specifications (reactor and detector locations, average efficiencies and running time as described in section~\ref{sec:comparison}).

Reactor neutrino experiments measure the survival probability $P_{ee}$ which does not depend on the Dirac $\delta_{\text{CP}}$ phase. In addition, the oscillation of MeV's reactor neutrinos studied over a distance of a few kilometers is not affected by the modification of the coherent forward scattering from matter electrons~\cite{minakatareactor2002,Lindnerreactor}. Expanding the full three~flavors \nueb oscillation probability as a function of $(\Delta{m}^2_{21}/\Delta{m}^2_{31})^2\simeq 1/30^2$ ratio and \ssqtt, $P_{ee}$ measurements from reactor experiments on the kilometer scale may be described by the simplest two~flavors oscillation formula:
\begin{equation}
 \label{eq:2nuSP}
 P_{ee} \simeq 1-\ssqtt\sin^2\left(\frac{\Delta{m}^2_{31}L}{4E}\right)\;.
\end{equation}
as long as $\ssqtt\gtrsim 10^{-3}$. We assume that in eq.~(\ref{eq:2nuSP}), $\Delta{m}^2_{31}$ is measured by other experiments (MINOS~\cite{MINOS}, K2K~\cite{K2K} and super-K~\cite{SKatm}).
With a determination of $\Delta{m}^2_{31}$ better than~10~\%, the impact on \ssqtt determination can be neglected~\cite{Mention}. All results within this study are, unless otherwise stated, computed for a representative value of $\Delta{m}_{31}^2$~\cite{schwetz}:
\begin{equation}
 \label{eq:def_dm2_31}
 \Delta{m}^2_{31} = 2.5_{-0.25}^{+0.25}\;10^{-3}~\text{eV}^2\;~\text{at 68~\% C.L.}.
\end{equation}

\section{Generic analysis of $\mathbf{\theta_{13}}$ sensitivity}
\label{sec:analysismetod}

The calculation of event rates is a convolution of the \nueb flux spectrum, the cross section, the oscillation probability, the detector efficiency with the energy response function.

The detector energy response simulation, as well as its correction through detector callibrations are specific to each experiment. The details of the corrections are thus beyond the scope of this article. We therefore assume in the following that the reconstructed energy is identical to the true deposited energy. We will implement, anyway, a simple energy scale systematic uncertainty (section~\ref{sec:systematics_review}).

We based our event rates computations on an extended version of the numerical code developed for Double~Chooz~\cite{Mention,DChooz}. These computations take into account the characteristics of each experimental setup as the number of reactors, detectors, locations, overburdens, efficiencies, operating time, and so on which will be described in section~\ref{sec:comparison}.

The resulting event rates, then, form the basis for a $\chi^2$ analysis, where systematic uncertainties are properly included. Since event rates in the disappearance channel of reactor experiments are quite large, we can use a Gaussian $\chi^2$, which has the advantage of allowing a natural inclusion of systematic errors through the so-called ``\textit{pull-approach}''~\cite{Fogli}:

\begin{equation}
  \label{eq:chi2}
  \chi^2 = \min_{\left\lbrace\alpha_k\right\rbrace} \left[
  \sum_{\begin{tabular}{>{$\scriptstyle}c<{$}}\\[-6mm]i=1,\ldots,\Nbins\\[-2mm] D=D_1,\ldots,D_{N_d}\end{tabular}}
  \left(\Delta_i^D-\sum_{k=1}^K \alpha_k S_{i,k}^D\right)^2 + \sum_{k,k'=1}^K\alpha_k W_{k,k'}^{-1} \alpha_{k'}
  \right]\;.
\end{equation}
This generic $\chi^2$ definition encompasses all the spectral information ($i$ index) from each detector ($D$ index) and systematics parameterization through $\alpha_k$ and $S_{i,k}^D$. For~$S_{i,k}^D=0$, we recover the classical $\chi^2$ definition, $\chi_{\text{no syst}}^2=\displaystyle\sum_{i,D}\left(\Delta_i^D\right)^2$, through
\begin{equation}
  \Delta_i^D=\left(N_i^{^{\star}D}-N_i^D\right)/U_i^D
  \label{eq:Delta_iD}
\end{equation}
where we assume that the simulated data event numbers, $N_i^{^{\star}D}$, are uncorrelated between bins and detectors. In the absence of real data, they are computed for fixed given values of $\Delta{m}_{31}^{2}{}^\star$ and $\ssqtt{}^\star$. On the other hand $N_i^D$, the theoretical model, relies on the searched \ssqtt value. We assume an uncorrelated weight error $U_i^D$ which, in the absence of systematic uncertainty, is simply expressed as the statistical error: $U_i^D=\displaystyle\sqrt{\smash[b]{N_i^D}}$.


Systematic uncertainties are included in the $\chi^2$ definition~(\ref{eq:chi2}) through $\alpha_k$ and $S_{i,k}^D$ coefficients.
The $S_{i,k}^D$ coefficient represents the shift of the $i^{\text{th}}$ bin of detector $D$ spectrum due to a $1\,\sigma$~variation in the $k^{\text{th}}$ systematic uncertainty parameter $\alpha_k$. Following this definition, we introduce bin, detector and reactor correlations in the systematic errors through $S_{i,k}^D$ definitions whereas systematic parameter correlations are gathered in $W_{k,k'}$ (we refer to the appendix for details). Eventually, some fully uncorrelated systematic uncertainties may be included through the $U_i^D$ definition, by adding quadratically all their effects together with the statistical uncertainty. We will use this property to include background shape uncertainties in our analysis as described in section~\ref{sec:backgrounds}. We refer to the appendix for the proper inclusion and definition of systematic coefficients $S_{i,k}^D$ and $\alpha_k$ inside the $\chi^2$ definition~(\ref{eq:chi2}).

We define the sensitivity or sensitivity limit $\ssqtt_{\text{lim}}$ as the largest value of \ssqtt which fits the true value $\ssqtt{}^{\star}=0$ at the chosen confidence level. We therefore determine the \ssqtt sensitivity at 90~\%~C.L. of an experiment as the value of \ssqtt for which
\begin{equation}
  \Delta\chi^2 = \chi^2(\ssqtt)-\chi^2_{\text{min}} = 2.71\;.
\end{equation}

\section{Generic overview of systematic error inputs}
\label{sec:systematic_inputs}

Systematic errors can be classified into three main categories: reactor, detector and data analysis induced  uncertainties. In this section we provide a brief and generic description of the systematic uncertainties included in our modelization. Details concerning specific experimental cases are given in sections~\ref{sec:comparison},~\ref{sec:sglexp_sensitivity}.

The dominant reactor induced systematic error comes from our limited knowledge of the physical processes which produce electron antineutrinos in nuclear reactors. This leads to an overall normalization error on the production rate of 1.9~\%~\cite{chooz}. Similarily we include a 2.0~\% uncertainty on the antineutrino spectral shape~\cite{nuspectrum}, with the conservative hypothesis that the energy bins are not correlated among themselves. Furthermore, even at a perfectly defined thermal power and with an absolute knowledge of the number of antineutrinos emitted for each fission, an underlying uncertainty remains since the nuclear energy released per fission is known to about 0.5~\%~\cite{chooz}. In our model the last three uncertainties are taken to be fully correlated between the nuclear cores.

We included another group of reactor induced systematics, taken to be uncorrelated between the reactor cores: the uncertainty on the determination of the thermal power of each nuclear core, within 0.6\,--\,3~\% and the uncertainty on the isotopic composition of the nuclear core fuel elements, within 2\,--\,3~\%. Also, finite size and solid angle effects, distances bias between reactors and detectors, as well as displacements of the neutrino production barycenter might affect the fluxes at the near detector(s) if they are close to the power plants (below 200~m) up to a level of 0.1~\%.

We did not implemented the uncertainty coming from neutrinos produced in the spent fuel pools, often located within tens of meters from the nuclear cores since we could not gather all the relevant information for the different sites.  We thus neglect this effect in our simulation, whithout any justification except in the case of Double Chooz, since a detailed evaluation has shown that this uncertainty does not affect its sensitivity~\cite{DChooz}. We point out that in particular cases this additional neutrino source slightly affects the antineutrino spectrum, and is thus relevant for experiments aiming at high sensitivities, \eg~$\ssqtt\sim 0.01$. We could also have introduced a specific error on the inverse-beta decay neutrino cross section of 0.1~\%~\cite{chooz,Eidelman:2004wy}. However, being fully correlated between the detectors, the latter can be gathered into a global uncertainty, adding up to the overall neutrino rate knowledge.

The basic principle of the multi-detector concept is the cancellation of the reactor induced systematics; additional contributions would not modify significantly the sensitivities of the forthcoming experiments. Let us now focus on the uncorrelated errors between detectors that could affect strongly the experimental results.

The uncorrelated errors between detectors directly contribute to the relative normalization of the measured antineutrino energy spectrum of each detector. One of the major improvement with respect to the CHOOZ experiment relies on the precise measurement of the number of free protons inside target volumes, proportional to the antineutrino rates. Experimentally the target mass will be determined at 0.2~\%. An uncertainty on the fraction of free hydrogen per unit volume remains, at the level of 0.2\,--\,0.8~\%, if different batches of liquid scintillator are used to fill the detectors of a given experiment. This complex case diserve a special treatment as described in section~\ref{sec:dayabay}. We did not include any error associated with the measurement of the live time of the experiments.

The last set of systematics concerns the selection cuts applied to extract the antineutrino signal and reject the backgrounds. Neutrino events are identified as positrons followed in time by a single neutron captured on a gadolinium nucleus. New detector designs have been proposed in order to simplify the analysis, reducing the systematic errors while keeping high statistics and high detection efficiency. We accounted for three uncertainties for both positron and neutron associated to a candidate event: the possibility of missing the particle as it escapes the target (escape), the uncertainty related to the particle interactions\footnote{Note that it affects essentially the neutrons, since the Gd concentration might differ between the detectors if they are not filled with a single batch of liquid scintillator.} (capture) and the identification cut based on the energy deposited in the detector active region (identification). Finally we take into account the uncertainty on the efficiency of the delay cut and an error on the neutron unicity of the event selections, whereas we do not consider any position vertex reconstruction. We provide in section~\ref{sec:systematics_review} (Table~\ref{tab:systematics_table}) the detailed systematic inputs necessary for our simulations. The uncertainty on these efficiencies have been treated, otherwise stated, as uncorrelated between detectors. Uncertainties induced by the background subtractions are discussed in sections~\ref{sec:backgrounds} and~\ref{sec:dsf}.

\section{Backgrounds: description and modelization}
\label{sec:backgrounds}

In this section we briefly review the three main kinds of background for the next generation of reactor neutrino experiments: accidental coincidences, fast neutrons and the long-lived muon induced isotopes $^9$Li/$^8$He. We then describe our simplified background modelization.


Naturally occurring radioactivity mostly creates accidental background, defined as a coincidence of a prompt energy deposition between 0.5 and 10~MeV, followed by a delayed neutron-like event in the fiducial volume of the detector within a few hundredths of a millisecond. Selection of high purity materials for detector construction (scintillator, mineral oil, PMT's, steel, \etc) and passive shielding provide an efficient handle against this type of background. We assume that the accidental background rate can be measured in situ with a precision of 10~\%.


Cosmic ray muons will be the dominating trigger rate at the depth of all near detector sites. Even though the energy deposition corresponds to about 2~MeV per centimeter path length (providing a strong discrimination tool) they induce the main source of background.


Muon induced production of the radioactive isotopes $^8$He, $^9$Li and $^{11}$Li can not be correlated to the primary muon interaction since their lifetimes are much longer than the characteristic time between two subsequent muon interactions. The characteristic signature of this last class of events consists in a four-fold coincidence ($\mu\to n\to\beta\to n$). The initial muon interaction is followed by the capture of spallation neutrons within about 1~ms. The time scale of the decay of the considered isotopes is on the order of a few 100~ms, again followed by a neutron capture. This background mimics the \nueb signal and is considered among the most serious difficulty to overcome for the next generation reactor neutrino experiments. In our simulation we will assume that it can be estimated to within 50~\%.

\begin{figure}[htpb]
\centering
\includegraphics[angle=-90,width=.5\textwidth]{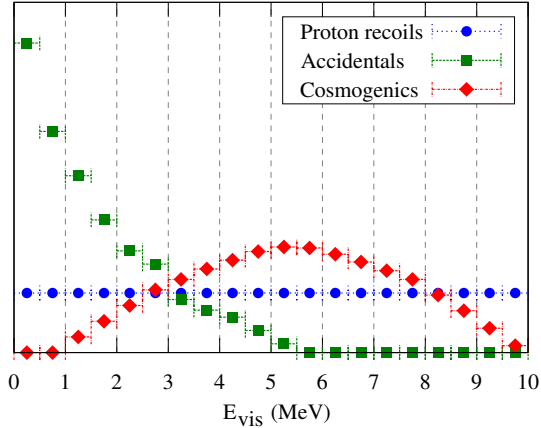}
\caption{Energy spectrum of the backgrounds from spallation neutrons, accidentals, and cosmogenic 20~\%~$^8$He and 80~\%~$^9$Li. Each curve is normalized to unity.}
\label{fig:backgrounds}
\end{figure}

A further source of background are neutrons that are produced in the surrounding rocks by radioactivity and in cosmic ray muon induced hadronic cascades. In the latter case, which is dominant at shallow depth, the primary cosmic ray muon may not penetrate the detector, being thus invisible. Fast neutrons may then enter the detector and create recoil protons and be captured by hydrogen or gadolinium nuclei after thermalization. Such a sequence can mimick a \nueb event. In the case of Double~Chooz far detector (depth of 300~\mwe), muon induced neutron production can be fairly well estimated from the results of the CHOOZ experiment, since it was the dominating background monitored during reactor off periods~\cite{chooz}. We will assume that this background rate can be estimated within a factor of two.
Figure~\ref{fig:backgrounds} illustrates the background spectra that we implemented in our modelization. We used the CHOOZ reactor
off data~\cite{chooz} to estimate the fast neutron background spectral shape, measured to be flat. We used a simple approximate exponential shape for accidental backgrounds. Finally we implemented the spectra of $^8$He and $^9$Li based on nuclear data information; we weighted them in the ratio 0.2/0.8, respectively. We checked that slight modifications of the background shapes do not change the results of our simulations for \ssqtt sensitivities above 0.01.

Background rates are adjusted to match to the predictions at each detector location. Afterward they are subtracted from the total $E_{\text{vis}}$ spectrum in each detector. These three backgrounds, $B_n$, have rates and shapes known at $\sigma_{B_n^D}$ and $\sigma_{\text{shp},B_n^D}$, respectively. We take the conservative assumption that these shape uncertainties are fully uncorrelated between bins. Thus, the $\sigma_{\text{shp},B_n^D}$ contributions may be directly included inside $U_i^D$ (eq.~(\ref{eq:Delta_iD})). These generic uncorrelated errors will then take into account the statistical and the background shape subtraction uncertainties:

\begin{equation}
 U_i^D=\left[
         N_i^D+\sum_{n=1}^{N_{\text{bkg}}} \left( B_{n,i}^D+\sigma_{\text{shp},B_{n}^D}^2\left(B_{n,i}^D\right)^2
         \right)
       \right]^{1/2}
 \label{eq:UiD}
\end{equation}
As the background rate uncertainties are correlated between bins, we treat them with additional pull terms, $\alpha_{B_n^D}$, in eq.~(\ref{eq:chi2}), with weights~$\sigma_{B_n^D}$ included in $S_{i,k}^D$ (see the appendix).

\section{Comparison of the current proposals}
\label{sec:comparison}

Several sites are currently being considered for new reactor experiments to search for~$\theta_{13}$: Angra dos Reis (Angra, Brazil), Chooz (Double~Chooz, France, and possibly Triple~Chooz), Daya~Bay (Daya~Bay, China), Kashiwazaki (KASKA, Japan) and  Yonggwang (RENO, Korea). All these experiments may be classified in two generations. The first aims to probe the value of \ssqtt till 0.02\,--\,03, and the second to track \ssqtt down to~0.01 (90~\%~C.L.). The first phase concerns Double~Chooz, RENO, and possibly Daya~Bay (with its phase~I). This phase should end by~2013. Angra, Daya~Bay (nominal setup with 8 detectors), KASKA and possibly Triple~Chooz are focusing on the second phase. For these second generation experiments, a significant R\&D effort is required since the effective Gd-scintillator mass will be increased by, at least, one order of magnitude, and systematics, as well as backgrounds uncertainties, have to be further reduced with respect to the first phase experiments. In the following comparisons we will not include Angra, KASKA and Triple~Chooz. We will only focus on Double~Chooz, Daya~Bay and RENO.

In the next sections we first introduce a generic discussion on systematics, scintillator composition and backgrounds (sections~\ref{sec:systematics_review} to~\ref{sec:dsf}). We will then shortly describe each setup and compute the associated baseline sensitivity (sections~\ref{sec:doublechooz} to~\ref{sec:reno}). We then perform a comparative analysis of the setups in section~\ref{sec:discusion} to highlight advantages and drawbacks of each setup.

\subsection{Detailed systematics review}
\label{sec:systematics_review}

The two Double~Chooz~\cite{DChooz} and Daya~Bay~\cite{DBProposal} proposals take a careful inventory of systematics, compiled in Table~\ref{tab:systematics_table} for comparison. Double Chooz and Daya Bay estimates for the case of no additional R\&D are at hand.
\begin{table}[htbp]
\fontsize{10}{12}\selectfont
\begin{center}
\scalebox{1}{
\begin{tabular}{l|c|cc|ccc}
\hline
\textbf{Error Description} & \textbf{CHOOZ} & \multicolumn{2}{c|}{\textbf{Double~Chooz}} & \multicolumn{3}{c}{\textbf{Daya~Bay}} \\
 &  &  &  &  & \textbf{No R\&D} & \textbf{R\&D} \\
 & \textbf{Absolute} & \textbf{Absolute} & \textbf{Relative} & \textbf{Absolute} & \textbf{Relative} & \textbf{Relative} \\
\hline
\hline
\textbf{Reactor} &  &  &  &  &  &  \\
\hline
Production cross section & 1.90~\% & 1.90~\% &  & 1.90~\% &  &  \\
Core powers & 0.70~\% & 2.00~\% &  & 2.00~\% &  &  \\
Energy per fission & 0.60~\% & 0.50~\% &  & 0.50~\% &  &  \\
Solid angle/Bary. displct. &  &  & 0.07~\% &  & 0.08~\% & 0.08~\% \\
\hline
\textbf{Detector} &  &  &  &  &  &  \\
\hline
Detection cross section & 0.30~\% & 0.10~\% &  & 0.10~\% &  &  \\
Target mass & 0.30~\% & 0.20~\% & 0.20~\% & 0.20~\% & 0.20~\% & 0.02~\% \\
Fiducial volume & 0.20~\% &  &  &  &  &  \\
Target free H fraction & 0.80~\% & 0.50~\% &  & ? & 0.20~\% & 0.10~\% \\
Dead time (electronics) & 0.25~\% &  &  &  &  &  \\
\hline
\textbf{Analysis (paticle id.)} &  &  &  &  &  &  \\
\hline
$e^+$ escape (D) & 0.10~\% &  &  &  &  &  \\
$e^+$ capture (C) &  &  &  &  &  &  \\
$e^+$ identification cut (E) & 0.80~\% & 0.10~\% & 0.10~\% &  &  &  \\
$n$ escape (D) & 0.10~\% &  &  &  &  &  \\
$n$ capture (\% Gd) (C) & 0.85~\% & 0.30~\% & 0.30~\% & 0.10~\% & 0.10~\% & 0.10~\% \\
$n$ identification cut (E) & 0.40~\% & 0.20~\% & 0.20~\% & 0.20~\% & 0.20~\% & 0.10~\% \\
\nueb time cut (T) & 0.40~\% & 0.10~\% & 0.10~\% & 0.10~\% & 0.10~\% & 0.03~\% \\
\nueb distance cut (D) & 0.30~\% &  &  &  &  &  \\
unicity ($n$ multiplicity) & 0.50~\% &  &  &  & 0.05~\% & 0.05~\% \\
\hline
\textbf{Total} & \textbf{2.72~\%} & \textbf{2.88~\%} & \textbf{0.44~\%} & \textbf{2.82~\%} & \textbf{0.39~\%} & \textbf{0.20~\%}
\\
\hline
\end{tabular}
}
\caption{Breakdown of the systematic errors included in the computation of
the sensitivity of Double~Chooz~\cite{DChooz} and Daya~Bay~\cite{DBProposal}. Since no breakdown of the RENO systematic errors has been published we use the same systematic error budget as for Double~Chooz. Double~Chooz and Daya~Bay relative systematics are almost comparable, the only main difference coming from the determination of the gadolinium concentration and the free proton fraction inside the target volume. The absolute determination of the free proton fraction can have some impact in Daya~Bay since multiple batches will be used to fill all the 8~detectors (see section~\ref{sec:dayabay} for details). Nevertheless, there is no published value in the Daya~Bay proposal~\cite{DBProposal}.}
\label{tab:systematics_table}
\end{center}
\end{table}
We decided to strictly use the systematic errors (Table~\ref{tab:systematics_table}) and background values quoted by the collaborations~\cite{DChooz,DBProposal,RENO}. However, we could not find any detailed background estimate for the RENO and Daya~Bay Mid site setups. In the latter cases we use a simple model to estimate the background subtraction uncertainties from the scaling of the Double~Chooz far detector (see section~\ref{sec:dsf}).

Through all the available publications~\cite{DChooz,DBProposal,RENO,KASKA}, the differences between the systematics are found only for the relative normalization of the dectetor~(\srel) and the subtraction uncertainties on background rates~(\sbkg). From section~\ref{sec:systematic_inputs} and Table~\ref{tab:systematics_table}, we thus group systematics in two categories to be used in our~$\chi^2$ analysis eq.~(\ref{eq:chi2}):
\begin{enumerate}
 \item generic systematics common to all the experiments (Table~\ref{tab:Chi2_syst}):
\begin{itemize}
 \item[--] \sabs, the theoretical uncertainty on reactor antineutrino spectrum prediction. We call it also the absolute normalization of event rates (since common to all the detectors), extracted from Table~\ref{tab:systematics_table} without power uncertainty contributions. \sabs is at the level of 2~\%;
 \item[--] \sshp, the theoretical reactor spectrum shape uncertainty, at the level of 2~\%;
 \item[--] \sscla, \ssclr, the absolute and relative energy scale uncertainties, roughly assessed at the level of 0.5~\% each;
 \item[--] \spwr, the reactor thermal power uncertainty, at the level of 2.0~\%;
 \item[--] \scmp, the reactor core specific fuel composition uncertainty which is roughly at the level of the power uncertainties (2\,--\,3~\%) on each fuel element.
\end{itemize}
\begin{table}[htpb]
\begin{center}
\begin{tabular}{*{6}c}
\hline
\sabs  & \sshp  & \sscla & \ssclr & \spwr  & \scmp \\
\hline\hline
2.0~\% & 2.0~\% & 0.5~\% & 0.5~\% & 2.0~\% & 2\,--\,3~\%\\
\hline
\end{tabular}
\caption{Generic systematic uncertainties as included in the $\chi^2$ analysis. For more details about the correlations between the systematics, we refer to the appendix, and more particularly to Table~\ref{tab:chi2paramtable}.}
\label{tab:Chi2_syst}
\end{center}
\end{table}

 \item specific systematics:
\begin{itemize}
 \item[--] \srel, the relative normalization of event rates between all the detectors. This uncertainty is uncorrelated between detectors;
 \item[--] \sbkg, the background subtraction unceratinties, described in section~\ref{sec:dsf}.
\end{itemize}
\end{enumerate}

\subsection{Impact of the scintillator composition}
\label{sec:scintillator}

In this section we stress the impact of the scintillator composition on the sensitivity of reactor neutrino experiments. All current projects will use a gadolinium doped liquid scintillator to enhance the neutron capture. The long-term stability of this scintillator is among the most difficult experimental challenges, since a degradation of the scintillator transparency would induce large systematic uncertainties. In the following we consider a stable scintillator for all experiments. The choice of the scintillating base has some importance since it defines the free proton number per unit volume, the \nueb rate and proton recoil background rate. Similarly the $^{12}$C number per volume drives the long-lived muon induced isotopes in the target scintillator.

\begin{table}[htpb]
\begin{center}
\scalebox{1}{
\begin{tabular}{cccccc}
\hline
Liquid				&	Formula		& density &	$10^{28}\,H/m^3$ & $10^{28}\,C/m^3$ \\
\hline
\hline
Dodecane (DOD)			& C$_{12}$H$_{26}$ 	& 0.753	  &	6.93	&	3.20	\\
pseudocumene (PC)			& C$_{9}$H$_{12}$ 	& 0.88	  &	5.30	&	3.97	\\
phenylxylylethane (PXE)		& C$_{16}$H$_{18}$ 	& 0.985	  &	5.08	&	4.52	\\
90~\% DOD+10~\% PC		& mixture in vol.	& 0.77	  &	6.77	&	3.32	\\
80~\% DOD+20~\% PXE		& mixture in vol.	& 0.80    &	6.56	&	3.46	\\
linear alkylbenzene (LAB)		& C$_{16}$H$_{30}$ 	& 0.86	  &	6.24	&	3.79	\\
\hline
\end{tabular}}
\caption{Impact of the scintillator composition on the target free proton number driving the \nueb rate as well as the recoil proton background, and on the carbon composition which drives the long-lived muon induced isotopes produced in the detector. In the following we consider that Double~Chooz is using modules containing 8.26~tons of
80~\% dodecane +20~\% PXE (in volume) based target scintillator, and that Daya~Bay and RENO are both using 20~ton modules of a LAB based target scintillator.}
\label{tab:scintillator}
\end{center}
\end{table}

Different bases can be used as neutrino target scintillator, mixture of dodecane (DOD), pseudocumene (PC) or phenylxylylethane (PXE), or linear alkylbenzene (LAB), as described in details in Table~\ref{tab:scintillator}. If we consider the Double~Chooz scintillator (80~\% DOD + 20~\% PXE) as our reference, a pure LAB scintillator contains 4.9~\% less free proton per volume, and 9.5~\% more carbon atoms. In the following we will assume the Daya~Bay~\cite{DBProposal} and RENO~\cite{RENO} experiments will use pure LAB as the target scintillator, and we will renormalize the neutrino and the backgound rates accordingly in section~\ref{sec:dayabay} and~\ref{sec:reno}.

\subsection{Estimation of the $\mu$-induced backgrounds}
\label{sec:dsf}

Our estimates of the backgrounds induced by cosmic muons is based on the Double~Chooz proposal~\cite{DChooz}. A modification of the MUSIC code~\cite{music} was used to compute the muon rates and energy spectra by propagating surface muons through rock. The site topographies have been included, according to a digitized map of the Chooz hill profile~\cite{tang} in the case of the far site (300~\mwe overburden), and according to a flat topography in the case of the near site (80~\mwe overburden). For the far site this full Monte-Carlo simulation predicts a muon flux $\Phi_{\mu}$=$0.612\pm 0.007~\text{m}^{-2}\,\text{s}^{-1}$, slightly higher than the approximate measured value quoted in~\cite{chooz}. The mean muon energy computed according to this method is $\left<E_{\mu}\right>=61$~GeV. For the near site, we get $\Phi_{\mu}=5.9~\text{m}^{-2}\,\text{s}^{-1}$, and $\left<E_{\mu}\right>=22$~GeV. A similar detailed computation was performed for the three sites of the Daya~Bay collaboration~\cite{DBProposal}.

Thus, for both the cases of Double~Chooz and Daya~Bay we use only the values of the muon flux and mean energy computed by the collaborations. However, we could not find any published data for the case of RENO. We then use the underground muon fluxes and mean energies calculated analytically following~\cite{bernhard}. In order to justify this approximation, Table~\ref{tab:depthscalingfactor} reports the muon flux and mean energy computed by the Double~Chooz and the Daya~Bay collaborations, from 300~\mwe to 923~\mwe, as well as the analytical computations according to~\cite{bernhard}. We found a reasonable agreement which bears out the use of the analytical model~\cite{bernhard} for the case of the RENO sites (depths of 255~\mwe and 675~\mwe). Nevertheless, note that the analytical simulation assumes a flat topography. It is worth noting, however, that the mean muon energy predicted by the analytical computation is systematically $\sim$\,25~\% lower in the depth range of interest. Therefore we arbitrarily renormalize the analytical calculation by 25~\% to estimate the mean muon energy at the RENO sites. We also note a large discrepancy between the detailed computation and the analytical model of~\cite{bernhard} for the Double~Chooz near site, probably due to its very shallow depth.

Cross sections of muon induced isotope production on liquid scintillator targets ($^{12}$C) have been measured by the NA54 experiment at the CERN SPS muon beam at 100~GeV and 190~GeV muon energies~\cite{NA54}. The energy dependence was found to scale as $\sigma_{\text{tot}}(E_{\mu})\propto \left<E_{\mu}\right>^{\alpha}$ with $\alpha = 0.73 \pm 0.10$ averaged over the various isotopes produced. We consider in the following that both long-lived muon induced isotopes and muon induced fast neutron backgrounds scale as $\Phi_{\mu}\times\left<E_{\mu}\right>^{\alpha}$, our reference being taken at the Chooz far site (full Monte-Carlo simulation). We define the depth scaling factor as
\begin{equation}
\text{DSF} = (\Phi_{\mu}\,\left<E_{\mu}\right>^{\alpha})/(\Phi_{\mu}\,\left<E_{\mu}\right>^{\alpha})_{\text{Double Chooz far}}\;,
\end{equation}
which is illustrated in the last column of Table~\ref{tab:depthscalingfactor} for various detector sites.
\begin{table}[htpb]
\begin{center}
\scalebox{1}{
\begin{tabular}{lc*6{r}}
\hline
Site	 & depth (\mwe), & \multicolumn{2}{c}{Detailed simulation} & \multicolumn{2}{c}{Analytical model} & DSF\\
	 & topography & $\Phi_{\mu}$	& $\left<E_{\mu}\right>$	& $\Phi_{\mu}$ 	& $\left<E_{\mu}\right>$&\\
	 &  & $\text{m}^{-2}\,\text{s}^{-1}$	& GeV 			& $\text{m}^{-2}\,\text{s}^{-1}$	& GeV\\
\hline
\hline
DC near	 & 80, flat & 5.9		& 22			& \hlc{\red 9.9}	& \hlc{\red 17}	& 6.80\\
RENO near& 230, hill& ---		& ---			& \hlc{\red 1.2}	& \hlc{\red 40}	& 1.57\\
DB near 1& 255, hill& \hlc{\red 1.2}    & \hlc{\red 55}		& 0.9		& 44			& 1.32\\
DB near 2& 291, hill& \hlc{\red 0.73}	& \hlc{\red 60}		& 0.72		& 49			& 1.06\\
DC far	 & 300, hill& \hlc{\red 0.61}	& \hlc{\red 61}		& 0.67		& 50			& 1 \\
DB mid	 & 541, hill& \hlc{\red 0.17}	& \hlc{\red 97}		& 0.15		& 71			& 0.32\\
RENO far & 675, hill& ---		& ---			& \hlc{\red 0.084}& \hlc{\red 94}	& 0.20\\
DB far	 & 923, hill& \hlc{\red 0.04}	& \hlc{\red 138}	& 0.035		& 118			& 0.10\\
\hline
\end{tabular}}
\caption{Muon flux $\Phi_{\mu}$ and mean energy $\left<E_{\mu}\right>$ for the underground site of the reactor
neutrino experiments. We compare the values obtained from a full Monte-Carlo simulation for Double~Chooz and
Daya~Bay to the analytical model of~\cite{bernhard}. We use the latter model for RENO. The depth scaling
factor (DSF) is defined by the product $(\Phi_{\mu} \times \left<E_{\mu}\right>^{\alpha})/(\Phi_{\mu} \times \left<E_{\mu}\right>^{\alpha})_{\text{Double \, Chooz \, far}}$, the Double~Chooz far site is taken as the reference. Backgrounds induced by cosmic muons are scaled according to this factor.}
\label{tab:depthscalingfactor}
\end{center}
\end{table}
Daily background rates computed for Daya~Bay, Double~Chooz, and RENO detectors at the different sites are then given in Table~\ref{tab:background_table}.
\begin{table}[htbp]
\fontsize{9}{11}\selectfont
\begin{center}
\scalebox{1}{
\begin{tabular}{@{\hspace{0mm}}c|c@{\hspace{1mm}}c|c@{\hspace{1mm}}c|c@{\hspace{1mm}}c@{\hspace{-6mm}}}
\hline
\textbf{Detector} & \multicolumn{2}{c|}{\textbf{Accidental (d$^{-1}$)}} & \multicolumn{2}{c|}{\textbf{$\mu$-induced fast-n (d$^{-1}$)}} & \multicolumn{2}{c}{\textbf{$\mu$-induced $^9$Li/$^8$He (d$^{-1}$)}}\\
\textbf{Site} & \textbf{Original} & \textbf{DC ext.} & \textbf{Original} & \textbf{DC ext.} &  \textbf{Original} & \textbf{DC ext.} \\
\hline
\hline
\textbf{Double~Chooz near} &     & $13.60\pm1.36$ &     & $1.36\pm1.36$ &     & $9.52\pm4.76$ \\
\textbf{RENO near} 	   & --- & $7.10\pm0.71$  & --- & $0.68\pm0.68$ & --- & $5.40\pm2.70$ \\
\textbf{Daya~Bay DB} 	   & $1.86\pm0.19$ & $5.98\pm0.60$ & $0.50\pm0.50$ & $0.57\pm0.57$ & $3.7\pm1.85$ & $4.55\pm2.27$ \\
\textbf{Daya~Bay LA} 	   & $1.52\pm0.15$ & $4.76\pm0.48$ & $0.35\pm0.35$ & $0.45\pm0.45$ & $2.5\pm1.25$ & $3.63\pm1.81$ \\
\textbf{Double~Chooz far}  & $2.00\pm0.20$  & --- & $0.20\pm0.20$ & --- & $1.40\pm0.70$ & --- \\
\textbf{Daya~Bay mid} 	   & --- & $1.45\pm0.14$  & --- & $0.14\pm0.14$ & --- & $1.10\pm0.57$ \\
\textbf{RENO far} 	   & --- & $0.90\pm0.09$  & --- & $0.09\pm0.09$ & --- & $0.69\pm0.35$\\
\textbf{Daya~Bay far} 	   & $0.12\pm0.01$ & $0.44\pm0.04$ & $0.03\pm0.03$& $0.04\pm0.04$ & $0.26\pm0.13$& $0.33\pm0.17$\\
\hline
\end{tabular}
}
\caption{Daily background rates computed for Daya~Bay, Double~Chooz, and RENO detectors at the different sites.
We consider the three main background sources: accidental events, $\mu$-induced fast neutrons, and $\mu$-induced $^9$Li/$^8$He. The columns labelled ``Original'' quote the background rates taken from the literature when available. The columns labelled ``DC ext.'' (for extended) quote the background value extrapolated from the background computed for the Double~Chooz far detector, scaled with the detector target mass, the scintillator free proton and carbon numbers, as well as the depth scaling factor (DSF).}
\label{tab:background_table}
\end{center}
\end{table}

\begin{table}[htbp]
\fontsize{9}{11}\selectfont
\begin{center}
\scalebox{1}{
\begin{tabular}{c|c@{\hspace{2mm}}c|c@{\hspace{2mm}}c|c@{\hspace{2mm}}c}
\hline
 & \multicolumn{6}{c}{\textbf{Background subtraction systematic errors (\%)}} \\
\textbf{Detector} & \multicolumn{2}{c|}{\textbf{Accidental}} & \multicolumn{2}{c|}{\textbf{$\mu$-induced fast-n}} & \multicolumn{2}{c}{\textbf{$\mu$-induced $^9$Li/$^8$He}}\\
\textbf{Site} & \textbf{Original} & \textbf{DC ext.} & \textbf{Original} & \textbf{DC ext.} &  \textbf{Original} & \textbf{DC ext.} \\
\hline
\hline
\textbf{Double~Chooz near} &     & $0.123$  &     & $0.123$ &     & $0.043$ \\
\textbf{RENO near} 	   & --- & $0.019$  & --- & $0.019$ & --- & $0.074$ \\
\textbf{Daya~Bay DB} 	   & $0.020$ & $0.064$ & $0.054$ & $0.061$ & $0.199$ & $0.245$ \\
\textbf{Daya~Bay LA} 	   & $0.020$ & $0.063$ & $0.046$ & $0.060$ & $0.164$ & $0.239$ \\
\textbf{Double~Chooz far}  & $0.292$ & --- & $0.292$ & --- & $1.020$ & --- \\
\textbf{Daya~Bay mid} 	   & --- & $0.120$  & --- & $0.115$ & --- & $0.458$ \\
\textbf{RENO far} 	   & --- & $0.100$  & --- & $0.095$ & --- & $0.382$\\
\textbf{Daya~Bay far} 	   & $0.010$ & $0.036$ & $0.025$ & $0.035$ & $0.108$& $0.138$\\
\hline
\end{tabular}
}
\caption{Background subtraction systematic errors (in percent) computed for Daya~Bay, Double~Chooz, and RENO detectors at the different sites. We consider the three main background sources: accidental events, $\mu$-induced fast neutrons, and $\mu$-induced $^9$Li/$^8$He. The columns labelled ``Original'' quote the systematic errors taken from the literature when available. The columns labelled ``DC ext.'' (for extended) quote the systematic errors extrapolated from the Double~Chooz far detector, taking into account the estimated detector signal to background ratio as well as the background rate uncertainty.}
\label{tab:background_subtraction_table}
\end{center}
\end{table}
Two cases are considered and compared: the background rates taken from the literature when available, and the background rates extrapolated from the background computed for the Double~Chooz far detector, scaled with the target mass, the scintillator free proton and carbon numbers, as well as the depth scaling factor (DSF).
We note here the good agreement between the Daya~Bay cosmogenic induced backgrounds ($^9$Li/$^8$He and fast neutrons) estimated from our Double~Chooz extended model and the original estimates of the Daya~Bay collaboration. This bears out the use of our model for the RENO and Daya~Bay mid site configurations. For the case of the $^9$Li/$^8$He background, we understand well this agreement since the background rate mainly depends on the mass of the neutrino target region. The agreement is more surprising, however, for the case of the fast neutron background, since the size of the liquid shielding around the detector active area is rather different between the Double~Chooz and Daya~Bay detectors (the RENO detector design is very close to the Double~Chooz case). In addition, further detector differences such as the thickness of the buffer oil shielding the inner target region, as well as the different mechanical structure explain the discrepancy between the Daya~Bay computation and the DC extended model for the case of the accidental background. Nevertheless, we found out that these differences influence only weakly the sensitivity computed for the three experiments since the accidental background energy spectrum  is different enough from the expected oscillation signal, and it is supposed to be known with a precision of 10~\%.

In a similar way, Table~\ref{tab:background_subtraction_table} gives the background subtraction systematic errors (in percent) computed for Daya~Bay, Double~Chooz, and RENO detectors at the different sites, taking into account the estimated detector signal to noise ratios as well as the background rate uncertainties.

\section{Reactor experiments baseline sensitivity}
\label{sec:sglexp_sensitivity}

\subsection{Double~Chooz}
\label{sec:doublechooz}

\begin{figure}[htpb]
\centering
\includegraphics[width=.5\textwidth]{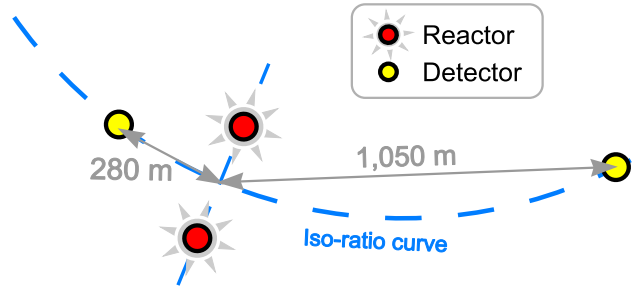}
\caption{Double~Chooz experiment site configuration. We show also on this figure the far flux iso-ratio line. Another detector located on this particular curve will receive the same reactor flux ratio as for the far detector: 44.5~\% from West and 55.5~\% from East reactor. The near detector of Double~Chooz is foreseen to be placed on this line.}
\label{fig:DoubleChooz_site}
\end{figure}

The Double~Chooz collaboration is composed of institutes from Brazil, France, Germany, Japan, Russia, Spain, United Kingdom, and the United States. The experimental site is located close to the twin reactor cores of the Chooz nuclear power station (two PWR\footnote{Pressurized Water Reactor} producing 8.5~GW$_{\textrm{th}}$), operated by the French company \'Electricit\'e de France (EDF).
\begin{table}[htpb]
\begin{center}
\begin{tabular}{lrr}
\hline
 \multicolumn{1}{c}{Detector}       &         near             &          far              \\
\hline
\hline
 Distance from West reactor  (m)    &       $290.7$            &   $1114.6\pm 0.1$         \\
 Distance from East reactor  (m)    &       $260.3$            &   $997.9\pm 0.1$          \\
 Detector Efficiency                &       80~\%              &   80~\%                   \\
 Dead Time                          &       25~\%              &   2.5~\%                  \\
 Rate without efficiency (d$^{-1}$) &       977                &   66                      \\
 Rate with detector efficiency (d$^{-1}$) & 782                &   53                      \\
 Integrated rate (y$^{-1}$)         &       $1.67\,10^{5}$     &   $1.48\,10^{4}$          \\
\hline
\end{tabular}
\caption{Double~Chooz antineutrinos rate expected in the near and far detectors, with and without reactor and detector efficiencies. The integrated rate in the last line includes detector efficiency, dead time, and reactor off periods averaged over a year. The averaged reactor global load factor is estimated at~79~\%~\cite{elecnuc}.}
\label{tab:setup_doublechooz}
\end{center}
\end{table}
The two, almost identical, detectors will contain a 8.3 ton fiducial volume of liquid scintillator (density of 0.8) doped with 0.1~g/l of gadolinium (Gd). The far detector will be installed in the existing laboratory, 1.05~km from the cores barycenter, shielded by 300~\mwe of rock. This detector should be operating alone for 1.5\,--\,2 years (DC~Phase~I), starting data taking by the end of~2008. The second detector will be installed in the meantime about 280~m from the nuclear core barycenter, at the bottom of a 40~m shaft (80~\mwe) to be excavated. Distances between detectors and nuclear cores as well as site overburdens are given in Table~\ref{tab:setup_doublechooz}. Since there are no more than two NPP cores, it is still possible to install the near detector at a suitable position where the ratio of reactor \nueb fluxes from each core is the same as for the far detector (the iso-ratio curve is plotted on figure~\ref{fig:DoubleChooz_site}).
This allows reactor relative uncertainty cancellations (NPP core compositions). This detector should be operational by~2010, and will take data for three years (DC~Phase~II). Other details concerning the experiment may be found in the collaboration proposals~\cite{DChooz}.

For the Double~Chooz phase~I (DC~Phase~I) analysis, we used systematics of Table~\ref{tab:Chi2_syst}, setting \srel and \ssclr to 0 since only one detector will be present. For a data taking period of 1.5~years, the sensitivity is $\ssqtt_{\text{lim}}=0.0544$. The second phase (DC~Phase~II) will then start and both detectors will take data for 3~years as scheduled in the proposal. The full experiment will then achieve a final sensitivity $\ssqtt_{\text{lim}}=0.0278$, assuming systematics from Table~\ref{tab:Chi2_syst} and $\srel=0.6~\%$~\cite{DChooz}.
The sensitivity worsens for $\Delta{m}_{31}^2<2.5\;10^{-3}~\text{eV}^2$, due to the close distance of the far detector\footnote{a bit too close to the NPP to get the maximum amount of information from first minimum of oscillation over the reactor spectrum distortion}. If we take the lower bound~\cite{limits,schwetz} on $\Delta{m}_{31}^2$ ($2.0\;10^{-3}~\text{eV}^2$), Double~Chooz will lose 30~\% in sensitivity whereas for the upper bound~\cite{limits,schwetz} on $\Delta{m}_{31}^2$ ($3.0\;10^{-3}~\text{eV}^2$), Double~Chooz will gain 15~\% in sensitivity.

\subsection{Daya~Bay}
\label{sec:dayabay}

\begin{figure}[htpb]
\hspace*{0cm}
\includegraphics[width=.4\textwidth,bb=14 14 397 451]{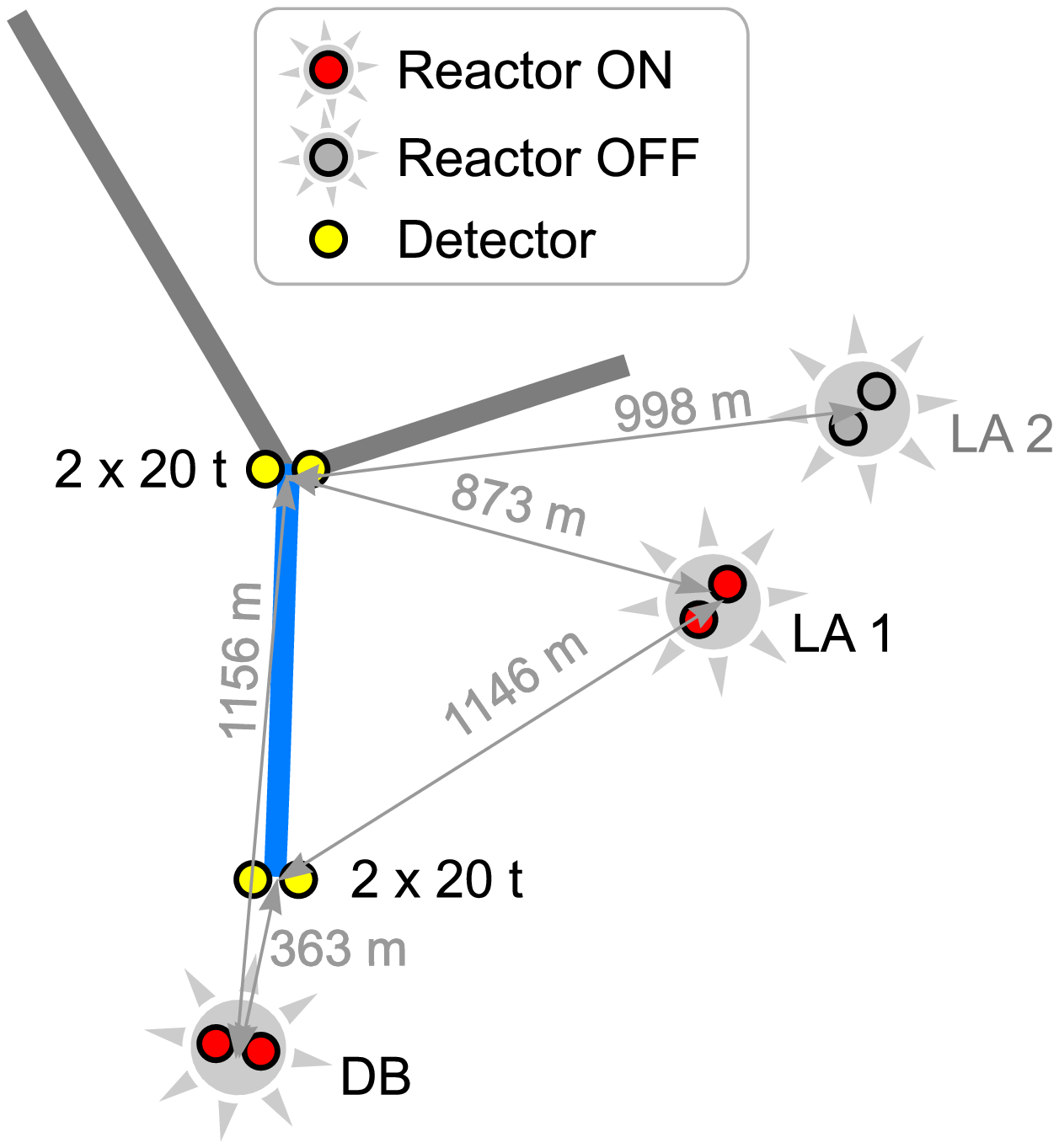}
\includegraphics[width=.4\textwidth,bb=14 14 397 451]{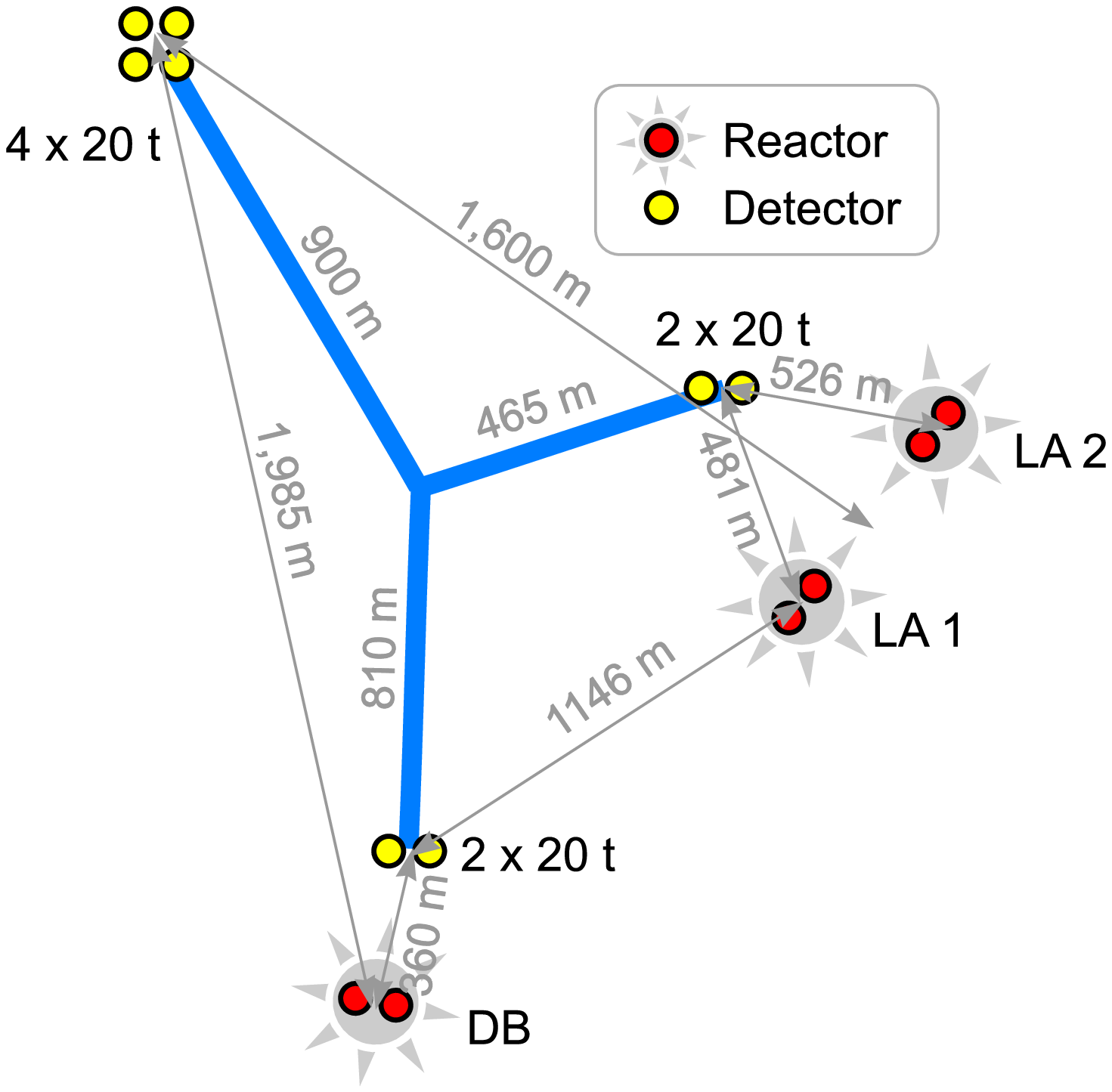}
\caption{Daya~Bay installation phases and site configuration. On the left: phase~I, with $2\times(2\times 20~\text{t})$ detectors, Daya~Bay and Ling Ao~I power plants are operating. On the right: phase~II with $2\times(2\times 20~\text{t})$ near sites, and $4\times 20~\text{t}$ at far site, all three power plants are operational.}
\label{fig:DB_site}
\end{figure}

Daya~Bay is an experiment proposed by institutes from China, the United States, and Russia. Daya~Bay~\cite{DBProposal} will be located in the Guang-Dong Province, on the site of the Daya~Bay nuclear power station. The site is made up of two pairs of twin reactors, Daya~Bay (DB) and Ling Ao~I (LA~I). An additional pair of reactors, Ling Ao~II (LA~II), is currently under construction and should be operational by~2010\,--\,2011~\cite{DBProposal}. Each core has a thermal power of 2.9~GW~\cite{elecnuc}. In the full installation setup 3.3~km of tunnel and 3~detector halls have to be excavated, in order to accommodate 8~detector modules~\cite{DBProposal}. Each module contains an effective volume of 20~tons of Gd-loaded LAB~liquid scintillator (Table~\ref{tab:scintillator}). Distances between detectors and nuclear cores as well as site overburdens are given in Table~\ref{tab:setup_dayabay}.
\begin{table}[htpb]
\begin{center}
\begin{tabular}{lrrrr}
\hline
 \multicolumn{1}{c}{Detector}&  near DB    &  near LA    &  mid   &    far        \\
\hline
\hline
 Distance from DB 1 (m)       &      350    &    1,356     &  1,153     & 1,970      \\
 Distance from DB 2 (m)       &      381    &    1,331     &  1,161     & 2,000      \\
 Distance from LA~I 1 (m)      &      942    &     492     &  783      & 1,619      \\
 Distance from LA~I 2 (m)      &     1,030    &     475     &  818      & 1,623      \\
 Distance from LA~II 1 (m)     &     1,378    &     500     &  968      & 1,602      \\
 Distance from LA~II 2 (m)     &     1,463    &     555     &  1,029     & 1,624      \\
 Detector eff.                &     80~\%   &    80~\%    &  80~\%    &  80~\%    \\
 Dead Time                    &     7.2~\%  &    4.3~\%   &   1~\%    & 0.2~\%    \\
 Rate without eff. (d$^{-1}$) &     1,938    &    1,813     &  494      &  430      \\
 Rate with detector eff. (d$^{-1}$) & 1,550  &    1,450     &  395      &  344      \\
 Integrated rate (y$^{-1}$)   & $4.10\,10^{5}$ & $3.95\,10^{5}$ & $1.11\,10^{5}$ & $9.77\,10^{4}$ \\
\hline
\end{tabular}
\caption{Daya~Bay antineutrino rates expected in the near and far detectors, with and without reactor and detector efficiencies. The integrated rate in the last line includes detector efficiency, dead time, and reactor off periods averaged over a year. We assumed that LA~II NPP will be operating for the time the far site will be fully installed, but for the Mid site installation we assumed that LA~II will be off.}\label{tab:setup_dayabay}
\end{center}
\end{table}
This site yields a rather complex signal composition in each detector coming from up to 6~different NPP cores, as shown in Table~\ref{tab:DB_rates}.
\begin{table}[htpb]
\begin{center}
\begin{tabular}{lccc}
\hline
& DB & LA1 & LA2 \\
\hline
\hline
near 1 (DB)    & 83.1~\% & 11.4~\% &  5.5~\% \\
near 2 (LA)    &  6.5~\% & 50.6~\% & 42.8~\% \\
Mid (LA2 OFF)  & 32.3~\% & 67.7~\% &  0.0~\% \\
Mid (LA2 ON)   & 22.5~\% & 47.1~\% & 30.4~\% \\
far            & 24.9~\% & 37.4~\% & 37.7~\% \\
\hline
\end{tabular}
\caption{Daya~Bay rate contributions from each NPP set (2 cores by set) while assuming, except if otherwise noticed (3$^{\text{rd}}$ line), the new Ling~Ao~II NPP is operating at full power.}\label{tab:DB_rates}
\end{center}
\end{table}
According to the Day~Bay proposal, our estimate of the sensitivity is $\ssqtt=0.0085$ very close to the Daya~Bay quoted value ($\ssqtt=0.008$). However, in the Daya~Bay proposal, the uncertainties on the reactor fuel composition, \scmp, as well as the energy scale associated uncertainties \sscla and \ssclr are neglected. Taking into account these systematics (Table~\ref{tab:Chi2_syst}) and the quoted value of $\srel=0.39~\%$ with no R\&D~\cite{DBProposal}, our estimate of the Daya~Bay final sensitivity is then $\ssqtt=0.009$, with the full installation (DB~Phase~II, Figure~\ref{fig:DB_site}) after 3~years of data taking. We draw the attention of the reader on the point that these computations are based on the assumption that $\srel=0.39~\%$ is fully uncorrelated between all the detectors, and in particular between detectors on a same site. Nevertheless this hypothesis is not guarenteed. We discuss this point as well as the current filling scenario (multi-batches) just below.

A preliminary fast phase (DB~Phase~I) is proposed by the Daya~Bay collaboration. This phase includes only 4~detectors, 2~of them located at the DB near site and the other~2 at the mid~site (see Figure~\ref{fig:DB_site}). Taking systematics from Table~\ref{tab:Chi2_syst} and $\srel=0.39~\%$ fully uncorrelated between detectors, we get a sensitivity after 1~year of data taking of \mbox{$\ssqtt=0.040$}. if LA~II NPP is off. If DB~Phase~I starts after~2010, with LA~II operational, we get a sensitivity of \mbox{$\ssqtt=0.038$} still after 1~year of data taking. The reason for a better sensitivity in the second scenario is that the mid detectors get 44~\% more \nueb~events (Table~\ref{tab:setup_dayabay}, \ref{tab:DB_rates} and Figure~\ref{fig:DB_site}), with a larger oscillation baseline with respect to LA~I.

\subsubsection*{Daya~Bay site correlation}

The main concept of the Daya~Bay experiment is based on the multi-inter-calibration of detectors. Since many detectors are installed on a same site, the total uncorrelated uncertainty of a site is decreased by a factor of $1/\sqrt{\smash[b]{N_d^S}}$ compared to the single detector uncorrelated uncertainty, where $N_d^S$ is the number of detectors on a given site. In the Daya~Bay proposal, the full relative uncertainty (\mbox{$\srel=0.39~\%$}) is assumed to be uncorrelated between all the detectors. However, the fraction of correlated and non-correlated error between detectors of a same site is not trivial. It relies on many experimental assumptions on uncertainty correlations. The absence of correlations between detectors on a same site implies there will be no detector-to-detector correction applied. If detector responses happen to be different, any data correction from detector to detector on a same site would yield correlations between detector uncertainties.

For DB~Phase~II, if we assume that \srel is fully uncorrelated between the detectors, we get a sensitivity of $\ssqtt=0.009$. On the contrary, if we assume that \srel is fully uncorrelated between detectors in different sites, but fully correlated between detectors on a same site, we get a sensitivity of $\ssqtt=0.012$. The real sensitivity should lay between these two extremes.

For DB~Phase~I, if we assume \srel is fully correlated on a same site and fully uncorrelated on distant sites, we get $\ssqtt=0.041$ if LA~II is off, and $\ssqtt=0.038$ if LA~II is on. We conlude from these results that DB~Mid sensitivity does not strongly depend on the correlations in \srel. More generally, DB~Mid setup weakly depends on \srel. This latter point will be discussed in section~\ref{sec:discusion}, together with a full description of the impact of each systematic on the forseen sensitivity.

\subsubsection*{Daya~Bay and the filling procedure}
\label{sec:filling}

\begin{figure}[htpb]
\centering
\includegraphics[angle=-90,width=.4\textwidth]{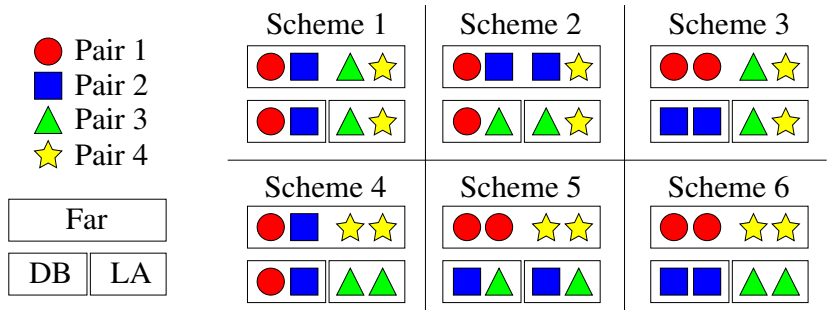}
\caption{Possible installation of detector pairs in the Daya~Bay experiment according to the adopted filling procedure~\cite{DBProposal}. Due to the large number of detectors, and the large amount of liquid to be managed, the Daya~Bay collaboration plans to fill detectors by pairs with four different Gd-doped liquid scintillator batches. The forseen detector installation scenario~\cite{DBProposal} corresponds to scheme~1. We illustrate here other installation possibilities (schemes 2~to~6).}
 \label{fig:filling_schemes}
\end{figure}

A large amount of Gd-loaded liquid scintillator will have to be produced, stored and filled into the detectors (8$\times$20~tons, which is 10~times more than in Double~Chooz). Due to the large number of detectors and the large amount of liquid to manage, the Daya~Bay collaboration plans to fill detectors with four different Gd-doped liquid scintillator batches. A single batch will be used to fill detectors by pairs~\cite{DBProposal}~(Figure~\ref{fig:filling_schemes}). The best installation scenario (which is the one chosen by the Daya~Bay collaboration) is then to move one of the filled detector to a near site, the other one to the far site (scheme~1 of~Figure~\ref{fig:filling_schemes} and Table~\ref{tab:chi2_filling_paramtable}). Another batch of Gd-loaded liquid scintillator will be used for the next pair of detectors and so on. With the adopted filling procedure~\cite{DBProposal}, extra systematic uncertainties on the hydrogen content between different batches have to be included.
\begin{figure}[htpb]
\centering
\includegraphics[angle=-90,width=.8\textwidth]{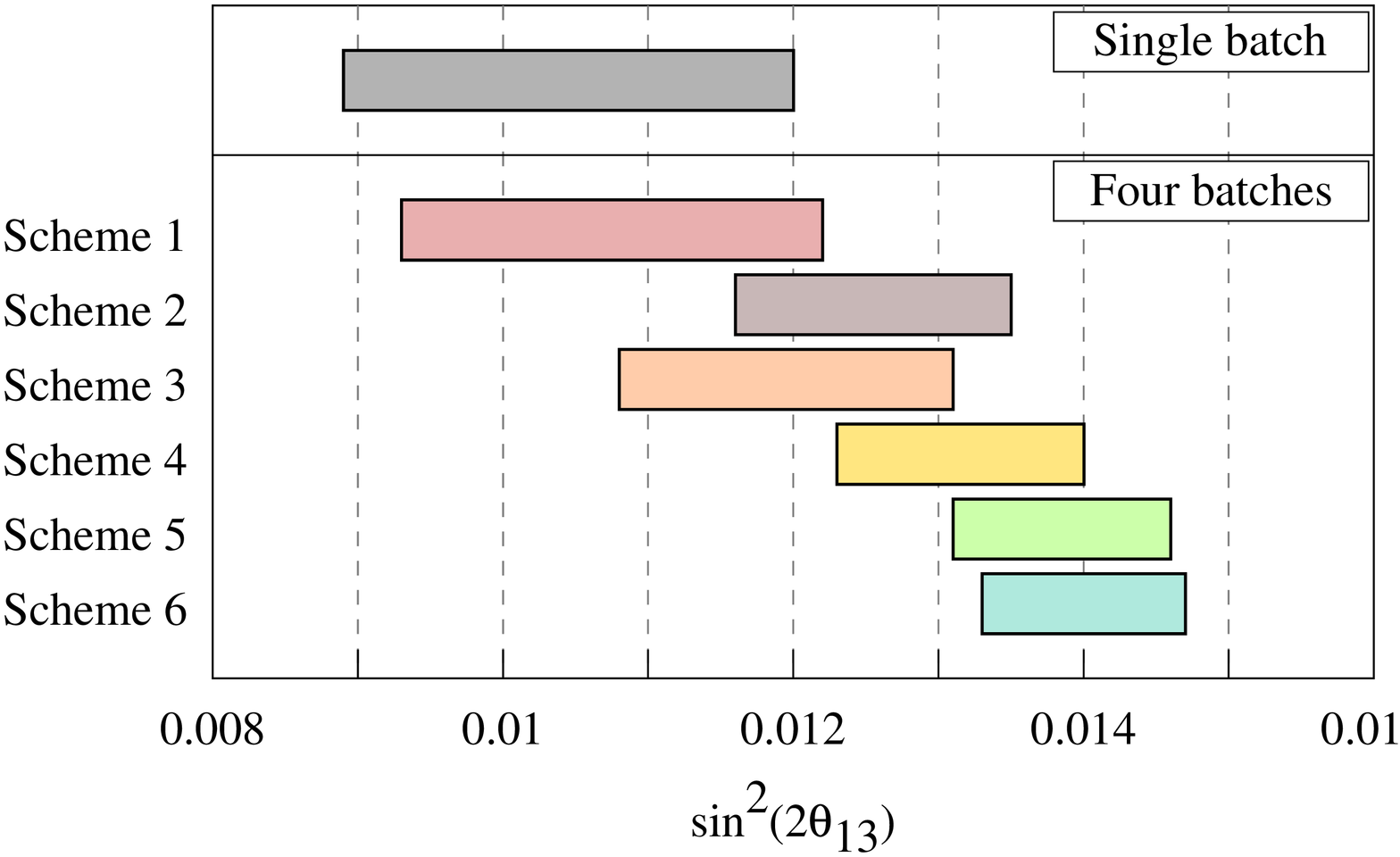}
\caption{Sensitivity on \ssqtt at~90~\% C.L. after 3~years of data taking for the 6~different installation schemes illustrated on Figure~\ref{fig:filling_schemes}, with \mbox{$\sptp=0.5~\%$}. Left bounds are computed assuming uncorrelated errors between all the detectors and right bounds are for the assumption of full correlation of \srel between detectors of a same site. The real sensitivity should be somewhere in between these two bounds. For comparison, we also show on this graph the computations for the single batch hypothesis where we take $\srel=0.39~\%$. Note that obviously the first installation scheme provides the best sensitivity. In these results we do not include any detector swapping scenario.}
\label{fig:Filling_sensitivity}
\end{figure}

Although the relative uncertainty on the free proton content within a same batch could be kept at a very low level (0.2~\% in the Daya~Bay proposal~\cite{DBProposal}, negligeable in the Double~Chooz proposal~\cite{DChooz}), it is not necessarily true in the case of different batches, in which the chemical composition may slightly change. The free proton content between different batches relies then on the measurement of this quantity. In the CHOOZ experiment~\cite{chooz}, the free proton fraction inside the Gd-loaded liquid scintillator was known to 0.8~\%. In Double~Chooz, this uncertainty is assessed at the level of 0.5~\%. Since there is no published value on the absolute determination of this quantity in the Daya~Bay proposal, we assume here, as in Table~\ref{tab:systematics_table}, a 0.5~\% uncertainty between different batches. The filling systematic coefficients are explained in Table~\ref{tab:chi2_filling_paramtable} (refer to the appendix and to Table~\ref{tab:chi2paramtable} for full description).
\begin{table}[htpb]
\fontsize{8}{8}\selectfont
\begin{center}
\scalebox{1}{
\begin{tabular}[t]{@{\hspace{3mm}}l@{\hspace{5mm}}c@{\hspace{5mm}}c@{\hspace{5mm}}c@{\hspace{5mm}}c@{\hspace{5mm}}}
  \hline\\[-2mm]
  {\bf Error type} & $\mathbf{k}$ & $\mathbf{S_{i,k}^D\times U_i^D}$ & $\mathbf{\alpha_{i,k}^D}$\\[2mm]
  \hline
  \hline
  \\[-2mm]
  \multicolumn{4}{@{\hspace{3mm}}l}{Filling ($\Ns=5\Ndet+\Nbins+5\Nreac+2$)}\\
  \hspace{.08\textwidth}of batch 1     & \Ns+1 & $\sptp \left(N_i^{D_1}+N_i^{D_5}\right)$ & $\alpha_{\text{pair},1}$\\
  \hspace{.08\textwidth}of batch 2     & \Ns+2 & $\sptp \left(N_i^{D_2}+N_i^{D_6}\right)$ & $\alpha_{\text{pair},2}$\\
  \hspace{.08\textwidth}of batch 3     & \Ns+3 & $\sptp \left(N_i^{D_3}+N_i^{D_7}\right)$ & $\alpha_{\text{pair},3}$\\
  \hspace{.08\textwidth}of batch 4     & \Ns+4 & $\sptp \left(N_i^{D_5}+N_i^{D_8}\right)$ & $\alpha_{\text{pair},4}$\\[2mm]
  \hline
\end{tabular}
}
\caption{Daya~Bay, phase~II (8 detectors), specific filling systematic parameters table in addition to standard one (see the appendix for details). Here we adopt a fully uncorrelated uncertainty of the different batches, $\sptp=0.5~\%$, but fully correlated in a same batch.}
\label{tab:chi2_filling_paramtable}
\end{center}
\end{table}
The uncertainty on the free proton fraction of a single batch is taken to $\sptp=0.5~\%$. This uncertainty is taken to be fully correlated when detectors are filled with the same batch and fully uncorrelated otherwise.

According to this filling procedure, the final sensitivity of DB~Phase~II would be $\ssqtt=0.0093$ instead of $0.0089$ with the initial installation scenario (scheme~1 of Figure~\ref{fig:filling_schemes}). In all the other configuration \mbox{schemes~(2\,--\,6)}, which allows comparing on a same site at least two detectors filled with the same batch, the sensitivity is more largely weakened (Figure~\ref{fig:Filling_sensitivity}), to an extent depending on~\sptp.

Note that we did not include any detector swapping option in previous conclusions. In the Day~Bay proposal, the retained installation scenario is the first scheme of Figure~\ref{fig:filling_schemes}. The baseline swapping option is then the permutation of two detectors filled with the same batch, in 4~steps, 1~step per batch. On the one hand, the drawback of such a swapping scenario is that two detectors filled with the same batch will never be directly compared. On the other hand, configuration \mbox{schemes~2\,--\,6} allow detector intercalibration within a pair. However, it should be noticed that the time spent in any configuration different from scheme~1 may decrease the combined final \ssqtt sensitivity (Figure~\ref{fig:Filling_sensitivity}).

\subsection{RENO}
\label{sec:reno}
\begin{figure}[htpb]
\centering
\includegraphics[width=.4\textwidth]{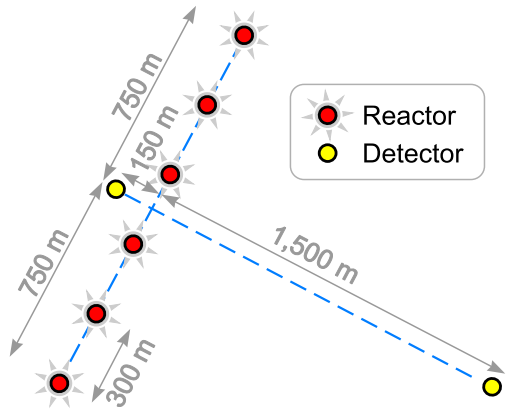}
\caption{RENO experiment site configuration.}
\label{fig:RENO_site}
\end{figure}

The RENO experiment~\cite{RENO} will be located close to the Yonggwang nuclear power plant in Korea, about 400~km south of Seoul. The power plant is a complex of six PWR reactors, each of them producing a thermal power of 2.74~GW~\cite{elecnuc}. The Yonggwong power station is ranked number~4 in the world, with a total thermal power of 16.4~GW. Its power rating is often cited as an advantage of the RENO experiment. These six reactors are equally distributed on a straight segment spanning over~1.5~km. The average cumulative operating factors for the reactors are all above 80~\%. Figure~\ref{fig:RENO_site} shows the foreseen layout of the experimental site.
The near and far detectors will be located 150~m and 1,500~m away from the center of the reactor row (Table~\ref{tab:setup_reno}), and will be shielded by a 88~m hill (230~\mwe) and a 260~m ``mountain'' (675~\mwe) respectively.
\begin{table}[htb]
\centering
\begin{tabular}{lrr}
\hline
 \multicolumn{1}{c}{Detector}             &     near     &     far    \\
\hline
\hline
 Distance from R1  (m)                    &      765     &     1677   \\
 Distance from R2  (m)                    &      474     &     1566   \\
 Distance from R3  (m)                    &      212     &     1507   \\
 Distance from R4  (m)                    &      212     &     1507   \\
 Distance from R5  (m)                    &      474     &     1566   \\
 Distance from R6  (m)                    &      765     &     1677   \\
 Detector Efficiency                      &      80~\%   &     80~\%  \\
 Dead Time                                &     7.2~\%   &    0.5~\%  \\
 Rate without efficiency (d$^{-1}$)       &     2859     &    121     \\
 Rate with detector efficiency (d$^{-1}$) &     2287     &     97     \\
 Integrated rate (y$^{-1}$)               & $6.20\,10^{5}$ & $2.82\,10^{4}$ \\
\hline
\end{tabular}
\caption{RENO antineutrino rates expected in the near and far detectors, with and without reactor and detector efficiencies. The integrated rate in the last line includes detector efficiency, dead time, and reactor off periods averaged over a year.}
\label{tab:setup_reno}
\end{table}
Two neutrino laboratories have to be excavated and equipped in order to host the detectors. They will be located at the bottom of two tunnels having a length of 100~m and 600~m for near and far detector, respectively. In this configuration, the flux contribution from $R_1$ to $R_6$ ranges from 3~\% to 39~\% for the near detector, and from 15~\% to 18~\% for the far detector (Table~\ref{tab:RENO_rates}).
\begin{table}[htbp]
\centering
\begin{tabular}{l*{6}{>{$}c<{$}}}
\hline
& R_1 & R_2 & R_3 & R_4 & R_5 & R_6\\
\hline
\hline
near & 3.0~\% & 7.8~\% & 39.2~\% & 39.2~\% & 7.8~\% & 3.0~\% \\
far & 14.8~\% & 16.9~\% & 18.3~\% & 18.3~\% & 16.9~\% & 14.8~\% \\
\hline
\end{tabular}
\caption{RENO rate contributions from each reactor core.}
\label{tab:RENO_rates}
\end{table}

Assuming the sytematics of Table~\ref{tab:systematics_table} and, lacking of published data, fixing $\srel=0.6~\%$ as for Double~Chooz, we obtain a final sensitivity after 3~years of data taking of $\ssqtt=0.021$. This is calculated for $2\times 20$~t detectors. However, a recent talk~\cite{RENO_NOW06}, quotes a different detector size, with 15~t of Gd-doped liquid scintillator. In that case, the final sensitivity after 3~years, would be $\ssqtt=0.023$.

As seen from Table~\ref{tab:RENO_rates}, the near detector monitors mainly the two central NPP cores, and the experiment sensitivity is quite affected by the reactor power uncertainties. If we compute the sensitivity with only the 2 central NPP cores, we get even better results compared to the full NPP configuration. This means that four of the six cores are useless for the \ssqtt measurement. This can be understood by the fact that the sensitivity gained by statistics is compensated by a loss of information on the \nueb rates and energy spectra. Thus the appeal of this site is diminished. The sensitivity is equivalent to a 5.8~GW$_{\text{th}}$ NPP reactor neutrino experiment for $\spwr=2.0~\%$.

The RENO collaboration considers using 3~small very near detectors (200\,--\,300~kg) to monitor sub-groups of cores of the NPP. However, taking into account current knowledge on reactor spectra ($\sabs=2.0~\%$, $\sshp=2.0~\%$~\cite{nuspectrum}), even dedicated detectors with $10^5$~\nueb events will not improve the thermal power knowledge below ${\sim}\,2~\%$. Thus, the overall \ssqtt sensitivity will not improve.

\section{Discussion}
\label{sec:discusion}

We develop here a two step comparison of the experiments described before: Double~Chooz phase~I (single detector, \textit{DC~I}), phase~II (both detectors, \textit{DC~II}), Daya~Bay phase~I (\textit{DB~Mid}) and phase~II (\textit{DB~Full}) and, RENO (\textit{RN}). The first elements of comparison are based on a single core equivalent approach. Although purely hypothetical, this analysis provides a lot of information on the impact of the layout of the site (locations of NPP reactor cores and detectors). The second approach, giving far more information on the impact of systematics on each experiment, is based on the \textit{pulls-approach}, presented in section~\ref{sec:analysismetod} and detailed in the Appendix. Note that in the following discussion we adopt the approximation that all the NPP cores operate with the same average efficiency. However, this assumption is not guaranteed, and, especially for NPP with many reactors such as RENO and Daya~Bay experiments, running time and procedure of each NPP core have to be taken into account in the final analysis. For the Double~Chooz experiment, which places the near detector on the flux iso-ratio line of the far detector, the full running operation time and procedure of each core is not needed to perform the final analysis.

\subsection{The single core equivalent approach}
\label{sec:SCE}

First of all we may simplify each experiment to its roughest single core equivalent (SCE) with total matching power $\overline{P}=\sum_{r=1}^{\Nreac} P^r$, with $P^r$ the power of the $r^{\text{th}}$ reactor and \Nreac the number of available reactors on site. In this case, we compute the \ssqtt sensitivity for each experiment (see Table~\ref{tab:sin_sensitivity_exp}) for their baseline option~\cite{DBProposal,DChooz,RENO} but also for a single core equivalent, and the averaged near and far detector locations computed with
\begin{equation}
\overline{L}=\left(\sum_{i=1}^{N_r} \frac{1}{N_r L_i^2}\right)^{-1/2}\;.
\end{equation}
This average distance, $\overline{L}$, yields the same event rates associated to near and far detectors of each experiment except for the Daya~Bay Phase~II experiment. For this particular case, we have to compute this average distance in two steps for the equivalent near detector. In this special case, since there are two near sites, we compute the $\overline{L}$ for the DB near site, $\overline{L}_{DB}$, and for the LA near site, $\overline{L}_{LA}$, and then compute the overall single near detector equivalent distance as
\begin{equation}
\overline{L}_{\text{N}}=\left(\frac{1}{3}{\overline{L}_{DB}}^{-2}+\frac{2}{3}{\overline{L}_{LA}}^{-2}\right)^{-1/2}\;.
\end{equation}
Following this SCE simplification, numerical values of \ssqtt sensitivity are gathered in Table~\ref{tab:sin_sensitivity_exp}.
\begin{table}[htpb]
\begin{center}
\begin{tabular}{|l|c|cc|c|c|}
\cline{2-6}
\multicolumn{1}{c|}{} & DC & \multicolumn{2}{c|}{DB Mid} & DB Full & RN\\
\multicolumn{1}{c|}{} &    & LA~II OFF & LA~II ON & &\\
\hline
$\cal L$ (in r.n.u.) & 17 & 67 & 101 & 303 & 71\\
$\overline{L}_{\text{far}}$ (in m) & 1,051 & 931 & 951 & 1,716 & 1579\\
$\overline{L}_{\text{near}}$ (in m) & 274 & 484 & 576 & 441$^\star$& 325\\
\hline
$\ssqtt_{\text{lim}}$ & 0.0278 & 0.0410 & 0.0381 & 0.0110 & 0.0213\\
$\ssqtt_{\text{lim}}^{\text{SCE}}$ & 0.0274 & 0.0274 & 0.0289 & 0.0105 & 0.0176\\
\hline
\end{tabular}
\caption{In this table we provide the luminosity (expressed in r.n.u., see eq.~(\ref{eq:rnu})), $\ssqtt_{\text{lim}}$ at 90~\%~C.L. of Double~Chooz (\textit{DC~II}), Daya~Bay phase~I (\textit{DB Mid}) and phase~II (\textit{DB Full}), and RENO (\textit{RN}). Also quoted in this table are the average distances, $\overline{L}$, to a single equivalent core (SCE) with total power $\overline{P}=\sum_{r=1}^{\Nreac}P^r$ (see text for explanations, the star ($^{\star}$) indicates a particular treatement). The sensitivity in the SCE case, $\ssqtt_{\text{lim}}^{\text{SCE}}$, is then computed to highlight possible drawback on the site configuration by comparison with the baseline sensitivity $\ssqtt_{\text{lim}}$.}
\label{tab:sin_sensitivity_exp}
\end{center}
\end{table}
As a first remark, the biggest discrepancy between $\ssqtt_{\text{lim}}$ and $\ssqtt_{\text{lim}}^{\text{SCE}}$ is as large as 40~\% for the DB~Mid experiment. With four to six~times higher luminosity compared to DC, a farther near site from the cores and a closer far site, the sensitivity for the hypothetical DB Mid SCE experiment is not improved with respect to DC. The discrepancy between DB~Mid and DB~Mid~SCE clearly comes from the wide repartition of NPP. LA~cores can not be properly monitored at the DB~near site. We conclude that the DB~Mid experiment is half way between an experiment with a single far detector and an experiment with two identical detectors, one near and one far. One may notice that DC is an experiment where the reactor cores may be considered as a single equivalent core with double power since there is only a less than 1~\% difference between DC and DC~SCE. This is not the case for the RN experiment where the relative discrepancy between RN and RN~SCE is at the level of 15~\%. Daya~Bay Phase~II, thanks to its two near sites, is only marginally affected by a $\sim$\,5~\% difference between DB Full and DB~Full~SCE (more on section~\ref{sec:pulls}).

\subsection{The pulls analysis}
\label{sec:pulls}

In this discussion we are interested in assessing how much a given experiment sensitivity relies on the knowldege of the systematics (\eg the number of systematics and the impact on sensitivity). The pulls-approach is perfectly adapted to this analysis. The idea is to break down the total $\Delta\chi^2$
\begin{eqnarray}
\label{eq:DeltaChi2}
 \Delta\chi^2(s_{\text{lim}}) &=& \chi^2(s_{\text{lim}})-\chi^2_{\text{min}} \\
  &=&  \min\limits_{\lbrace\alpha_1,\ldots,\alpha_K\rbrace}\chi^2(s_{\text{lim}},\alpha_1,\ldots,\alpha_K)-
       \min\limits_{\lbrace s,\alpha_1,\ldots,\alpha_K\rbrace}\chi^2(s,\alpha_1,\ldots,\alpha_K)\nonumber\;,\\
  & & \text{(where $s$ is shorthand for \ssqtt)}
\end{eqnarray}
into sub-parts $\delta\chi^2_i$ which represent their relative contribution to the overall $\Delta\chi^2$ of eq.~(\ref{eq:DeltaChi2}). Since the sensitivity limit on \ssqtt is computed at 90~\%~C.L., $\Delta\chi^2$ has a common value for every experiment and is equal to~2.71. Thus, $\delta\chi^2_i$ is defined as
\begin{equation}
 \delta\chi^2_i = \frac{\left(\text{i$^{\text{th}}$ pull term}\right)^2}{\Delta\chi^2}
\end{equation}
with the trivial induced normalization: $\sum_i \delta\chi^2_i=1$.

Table~\ref{tab:chi2_contrib} shows the results of our computation for the baseline option of each experimental setup.
We report, together with the final sensitivities discussed earlier, the relative contributions $\delta\chi^2_i$ where $\delta\chi^2_{N_1,N_2,F}$ are the \textit{observables} (the contribution from the first term in eq.~(\ref{eq:chi2}) for the respective detector) and the other $\delta\chi^2_i$ are the \textit{pulls} (the weight term contributions in eq.~(\ref{eq:chi2})).
\begin{table}[htpb]
\begin{center}
\begin{tabular}{|c|>{$}l<{$}|@{\hspace{6.5pt}}*{5}{c}|}
\cline{2-7}
\multicolumn{1}{c|}{}&\delta\chi^2_i\text{ in~\%} & DC~I & DC~II & DB~Mid & DB~Full & RN\\
\hline
\multirow{3}*{\rotatebox{90}{obs.}}& \delta\chi^2_{N_1} & --- & 3.0~\% & 4.3~\% & 1.1~\% & 1.5~\% \\
& \delta\chi^2_{N_2} & --- & --- & --- & 3.3~\% & --- \\
& \delta\chi^2_{F} & \hlc{\red 29.5~\%} & \hlc{\red 38.0~\%} & \hlc{\red 23.4~\%} & \hlc{\red 31.2~\%} & \hlc{\red 34.4~\%} \\
\hline\\[-14pt]
\multirow{9}*{\rotatebox{90}{pulls}}
& \delta\chi^2_{\text{abs}} & \hlc{\red 29.1~\%} & 1.5~\% & \hlc{\org 9.0~\%} & 1.0~\% & \hlc{\org 7.9~\%} \\
& \delta\chi^2_{\text{shp}} & \hlc{\org 18.4~\%} & 1.3~\% & \hlc{\org 8.4~\%} & 0.5~\% & 1.0~\% \\
& \delta\chi^2_{\text{rel}} & --- & \hlc{\red 48.3~\%} & \hlc{\org 6.6~\%} & \hlc{\red 56.8~\%} & \hlc{\red 28.5~\%} \\
& \delta\chi^2_{\text{scl,abs}} & \hlc{\org 6.5~\%} & 1.2~\% & \hlc{\org 6.1~\%} & 0.1~\% & 0.1~\% \\
& \delta\chi^2_{\text{scl,rel}} & --- & \hlc{\org 5.0~\%} & \hlc{\org 11.8~\%} & 1.6~\% & 0.2~\% \\
& \delta\chi^2_{\text{bkg}} & 1.0~\% & 0.8~\% & 0.4~\% & 0.1~\% & 0.5~\% \\
& \delta\chi^2_{\text{pwr}} & \hlc{\org 14.7~\%} & 0.8~\% & \hlc{\red 27.3~\%} & 3.9~\% & \hlc{\org 16.9~\%} \\
& \delta\chi^2_{\text{cmp}} & 0.1~\% & 0.0~\% & 0.6~\% & 0.1~\% & \hlc{\org 9.1~\%} \\
& \delta\chi^2_{\varepsilon} & 0.6~\% & 0.0~\% & 2.2~\% & 0.3~\% & 0.0~\% \\
\hline
\multicolumn{1}{c|}{}& \sin^2(2\theta_{13})_{\text{lim}} & 0.054 & 0.028 & 0.041 & 0.011 & 0.021\\
\cline{2-7}
\end{tabular}\caption{Relative contributions, $\delta\chi^2_i$, to the global $\Delta\chi^2$. The higher is the $\delta\chi^2_i$ contribution, the more the \ssqtt sensitivity depends on the considered parameter. In red we highlight the main contributions \mbox{(20$\,$--$\,$60~\%)}, in orange other significant terms \mbox{(5$\,$--$\,$20~\%)}. All the values are calculated for the base case of each experimental setup. For the DB~Full setup, where correlations between detectors on a same site are possible, we take \srel half correlated and half uncorrelated between detectors of a same site, and completely uncorrelated between detectors of different sites (see section~\ref{sec:filling} for details). Note that for Daya~Bay we do not include here batch-to-batch uncertainties.}
\label{tab:chi2_contrib}
\end{center}
\end{table}
Since the role of the far detector is to determine \ssqtt, it is obvious that the associated residual should significantly contribute to the sensitivity. However what we see is that all computed sensitivities mainly depend on systematics, which contribute from~60~\% to~70~\% of the overall $\Delta\chi^2$. In the concept of identical detectors, correlated uncertainties between detectors should weakly impact the sensitivity (at the level of near detector ``\textit{precision}''). This is automatically the case if the near detector successfully monitors the NPP cores. Two of the quoted experimental setups reach this goal: DC~II and DB~Full. Double~Chooz, with its final two detector installation, has one dominant systematic: the relative normalization uncertainty. The second most important contribution comes from the relative energy scale uncertainty.

Daya~Bay full installation is mostly limited by the relative normalization uncertainty. The weaker impact of the relative energy scale uncertainties comes from the better far site distance to the NPP~Cores. The uncertainty on the energy scale matches less the oscillation induced distortion. On the other hand, three of the described experimental setups still rely on theoretical knowledge of the spectrum and NPP~cores associated uncertainties: DC~I, DB~Mid and RN. In particular, the DB~Mid installation is sensitive to several systematics, especially the NPP power uncertainties. Taking data on a longer time scale (3~years, for instance) with such a detector configuration will not improve the $\ssqtt$ sensitivity as much. Single core power uncertainties do not weaken the sensitivity in two particular cases:
\begin{itemize}
\item[--] if the near and far detectors have the same NPP~core flux ratio contributions (this is the case of DC~II);
\item[--] if the far detector distance to each NPP~core is the same, even if the spectra ratios are not the same in near/far detectors. In this particular case, the oscillation pattern is not entangled with power uncertainties in the far detector, since the \nueb travelling distances are the same. Power uncertainties would only contribute, weakly, through the absolute normalization and correlations with other systematics in the near detector.
\end{itemize}
The DB~site configuration does not meet any of the above conditions. Moreover, for DB~Mid, no near site monitors the LA~I NPP. This makes the DB~Mid setup an intermediate between a two identical near/far detector experiment and a single far detector configuration. Since the near site is farther away from the NPP~cores (Table~\ref{tab:sin_sensitivity_exp}), theoretical uncertainties on the spectrum have a larger impact than in DC~II. Also, since the average distance of DB~Mid far site is closer, the oscillation pattern matches slightly better with the energy scale associated distortion (bigger contribution than in DC~II).

The RENO far site location is the best among the quoted setups in canceling the impact of the relative and absolute energy scale uncertainties. However, because the site configuration does not fill any of the two conditions for the cancellation of the NPP~core power uncertainties, this experiment relies on the precision with which each core power can be determined. Moreover, the near detector is a bit farther away than in the DC~II case, which explains why the global reactor \nueb rate is less effectively determined and have a larger contribution than in DC~II to the final \ssqtt sensitivity.

\subsection{Touching the ``right systematic chord''}
\label{sec:syst_string}

In the previous section we have determined the dominant systematics of each setup. In this section we focus on the comparison of all the reactor experiments under the assumption that systematics are known at the same level. Moreover, we want to illustrate the impact of the determination of the most significant systematics on the sensitivity of each experimental setup:
\begin{itemize}
\item the single core power uncertainties;
\item the relative normalization between detectors;
\item the energy scale uncertainties (absolute and relative, between detectors).
\end{itemize}
\begin{figure}[htpb]
\centering
\begin{tabular}{cc}
\includegraphics[width=.485\textwidth,angle=-90]{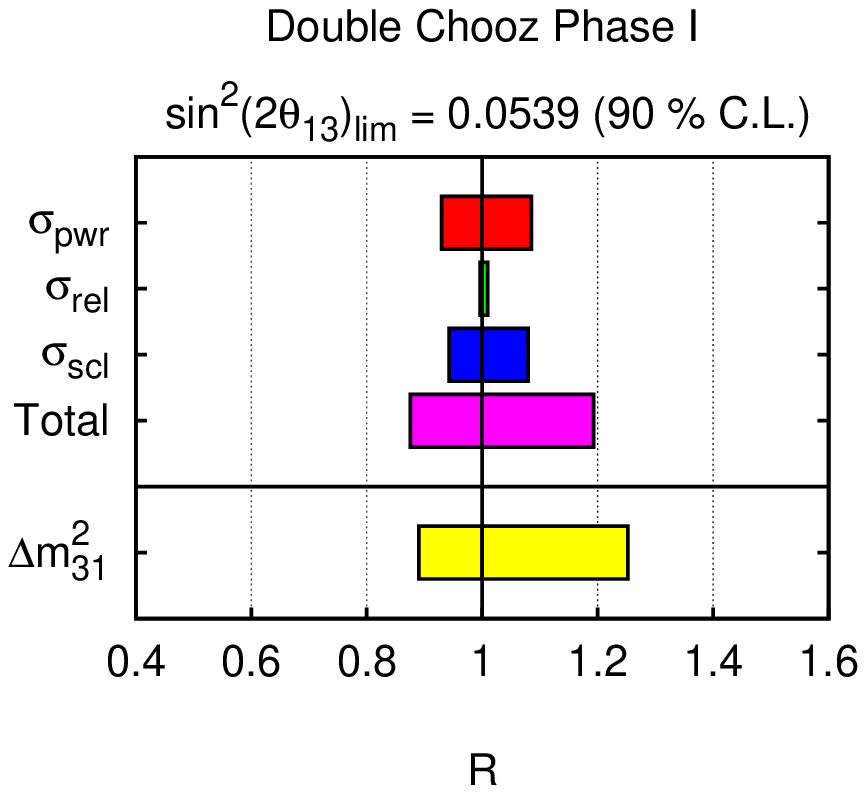}  &
\includegraphics[width=.485\textwidth,angle=-90]{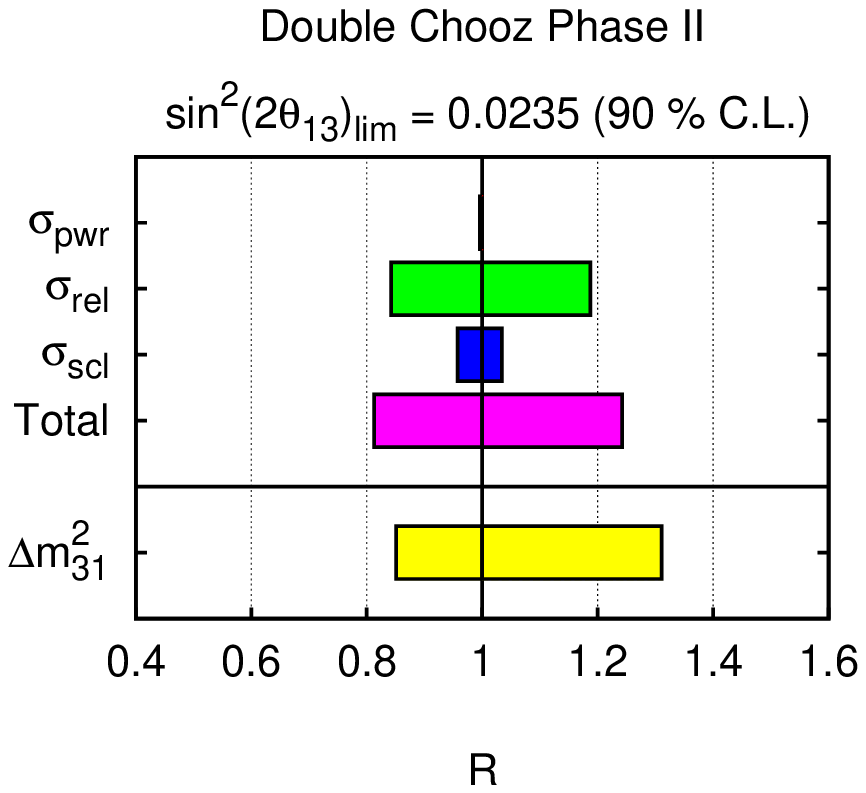} \\[-30mm]
\includegraphics[width=.485\textwidth,angle=-90]{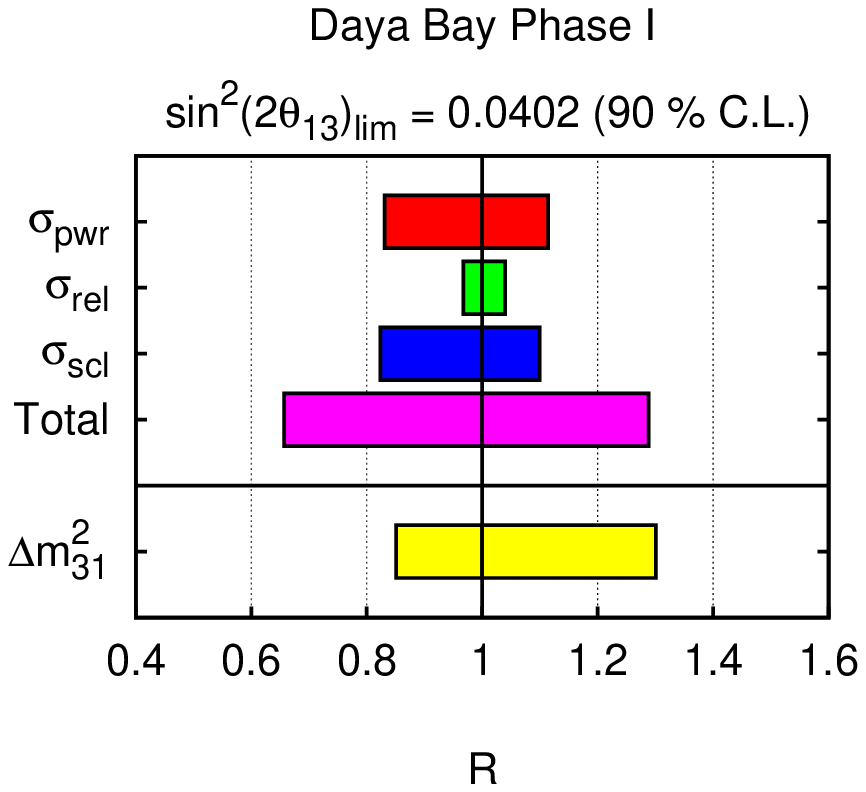}  &
\includegraphics[width=.485\textwidth,angle=-90]{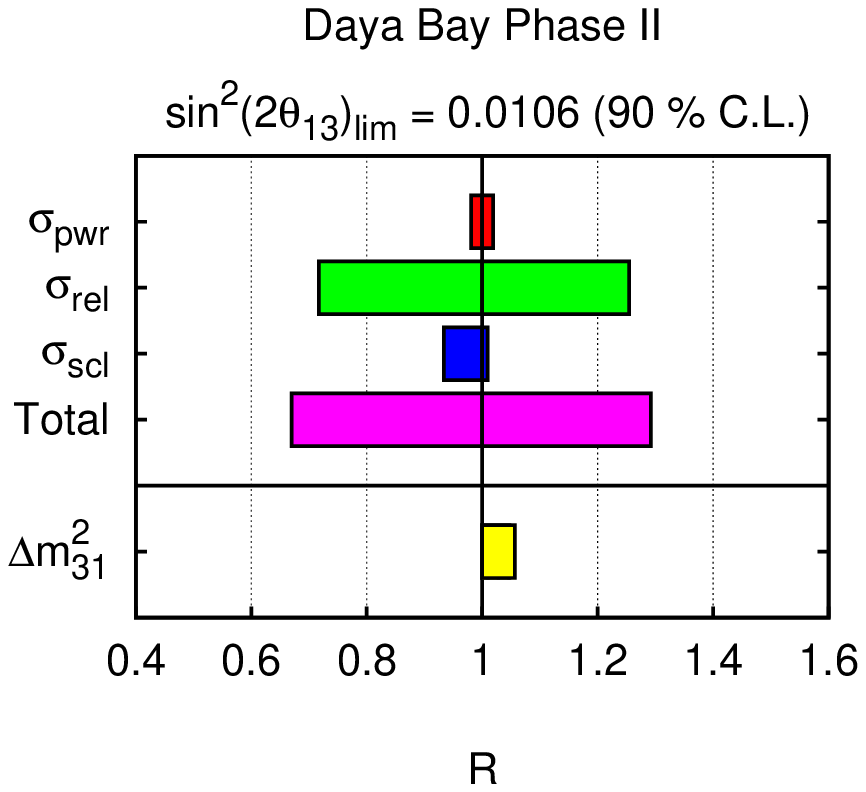} \\[-30mm]
\includegraphics[width=.485\textwidth,angle=-90]{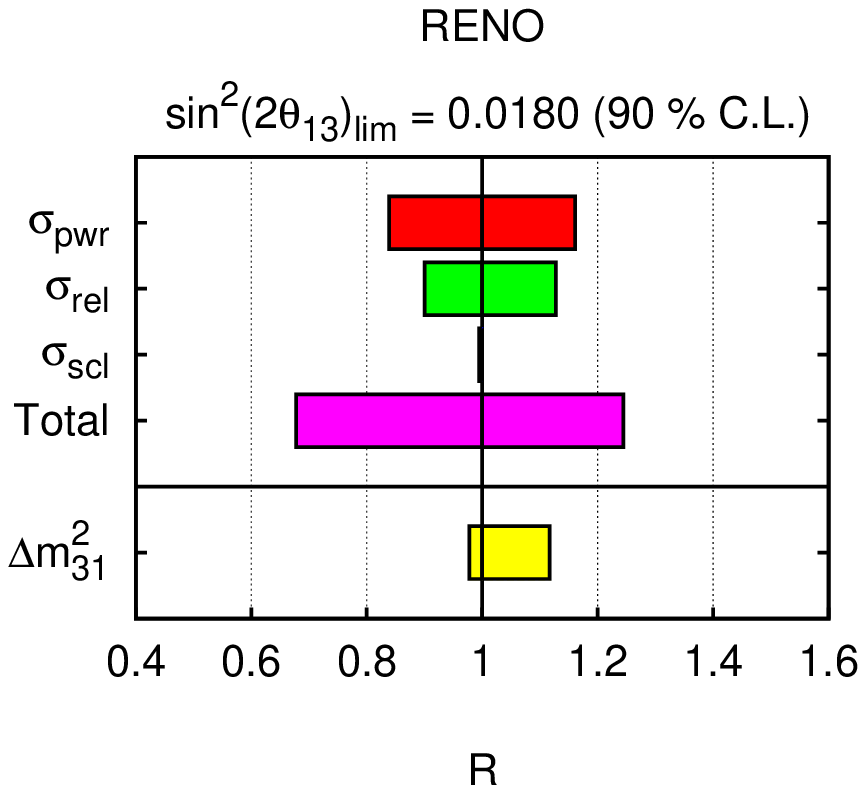}   &
\begin{minipage}[t]{.3\textwidth}
\vspace{12mm}
\fontsize{8}{10}\selectfont
\begin{flushright}
\begin{tabular}{|cc|}
\multicolumn{2}{c}{Common systematic framework}\\
\hline
\sabs & 2.0~\%  \\
\sshp & 2.0~\%  \\
\srel & 0.4~\% \\
\spwr & 2.0~\%  \\
\sscla\;\ssclr & 0.5~\% \\
$\Delta{m}^2_{31}~(\text{eV}^2)$ & $2.5\,10^{-3}$ \\
\hline
\end{tabular}\\
\vspace*{1mm}
\begin{tabular}{|l|c|c|}
\cline{2-3}
\multicolumn{1}{c|}{}& Best & Worst\\
\hline
\spwr & 0.6~\% & 3.0~\% \\
\srel & 0.2~\% & 0.6~\% \\
\sscla \ssclr & 0.0~\% & 1.0~\% \\
$\Delta{m}^2_{31}~(\text{eV}^2)$ & $3\,10^{-3}$ & $2\,10^{-3}$ \\
\hline
\end{tabular}
\end{flushright}
\end{minipage}
\end{tabular}
\\[-15mm]
\caption{Double~Chooz, Daya~Bay and RENO sensitivities as a function of the size of the main systematics.
The common systematic framework is what experimentalists believe to be achievable, without any further R\&D. It is worth noting that the main difference between the common and the baseline cases comes from Double~Chooz, which takes a conservative value of the relative normalization at 0.6~\%.
The common framework is used to compute the reference $\ssqtt$ sensitivity of each setup (value on top of each graph). Then each systematic (\spwr, \srel, \sscl) impact on sensitivity is separately computed and illustrated as ratio $R=\ssqtt_{\text{best or worst}}/\ssqtt_{\text{baseline}}$ on each graph. The overall impact changing all three systematics together is also illustrated with the ``Total'' label. Moreover we also provide a quick guess on \ssqtt sensitivity behaviour as a function of $\Delta{m}^2_{31}$ best fit value provided by other experiments. For the Daya~Bay~Phase~II experiment, where possible correlation between detectors on a same site may happen, we take \srel half correlated and half uncorrelated between detectors of a same site, and completely uncorrelated between detectors of different sites (see section~\ref{sec:filling} for details).}
 \label{fig:AllReacExpSensi}
\end{figure}
In the CHOOZ experiment, the power uncertainties were assessed at 0.6~\%~\cite{chooz}. However, this estimate was uniquely based on the heat balance of the steam generators. Even if quite precise, this method could be inaccurate. Other methods, such as the external neutron flux measurements, more directly linked to the fission rates inside the cores, lead to a power evaluation less effectively determined with an assessed error around 1.5~\%. This latter method allows continuous tracking of the NPP~core power variations. The Double~Chooz and Daya~Bay proposals set their baseline estimates of these uncertainties at a conservative value of 2~\%. We studied the impact of this knowledge on the sensitivity by assuming two extreme scenarios: in the worst case, a 3~\% error, and the best case of 0.6~\% precision. As a standard, we set central value $\spwr=2.0~\%$ for all the experiments.

The relative normalization between detectors is the most significant systematic in two identical detector setups, with a near detector successfully monitoring the whole NPP (DC~II and DB~Full setups). The two available detailed quantifications of this uncertainty are the Double~Chooz~\cite{DChooz} and Daya~Bay~\cite{DBProposal} proposals. Each detector is estimated to measure the \nueb rate to a relative accuracy with respect to each other of 0.39~\% (DB~Full) and 0.44~\% (DC~II). In the Double~Chooz proposal, however, a conservative value has been preferred for the baseline sensitivity calculation: $\srel=0.6~\%$. We will take this number as the worst case. The Daya~Bay collaboration plans, after some R\&D, to reach a relative uncertainty of 0.2~\%. This value has been taken as the best case. We set as a standard central value $\srel=0.4~\%$ for all experiments.

Even if we did not implement detector response in this simulations, we included energy scale uncertainties. The Double~Chooz proposal quotes $\sscla=0.5~\%$ and $\ssclr=0.5~\%$. This is taken as the common central values for all the experiments. However, in the CHOOZ experiment~\cite{chooz}, the energy scale uncertainty was estimated at the level of 1.1~\%. We thus take a 1.0~\% uncertainty on both relative and absolute energy scale determination as the worst case. As a best case, we switched off the impact of energy scale systematics on the sensitivity.

Although the impact of the $\Delta{m}^2_{31}$ uncertainty on the sensitivity to \ssqtt is negligible with the current knowledge, the central best fit value on this parameter from other experiments such as MINOS~\cite{MINOS}, K2K~\cite{K2K} and SuperK~\cite{SKatm}, may have some influence on the reactor experiment sensitivities. We thus include the current $\Delta{m}^2_{31}$ bounds on the sensitivity computations: the ``worst case'' is taken to be \mbox{$\Delta{m}^2_{31}=2.0\,10^{-3}~\text{eV}^2$} and the ``best'' one \mbox{$\Delta{m}^2_{31}=3.0\,10^{-3}~\text{eV}^2$} (2\;$\sigma$ bounds from~\cite{schwetz}). The standard central value for all the experiments has been taken to be \mbox{$\Delta{m}^2_{31}=2.5\,10^{-3}~\text{eV}^2$}.

We illustrate in Figure~\ref{fig:AllReacExpSensi} these three systematic scenarii: best, central and worst cases. Each contribution is assessed separately, but we also show on this graph the total impact by summing the effect of the three discussed systematics \mbox{(\spwr, \srel and \sscl)} to their respective best, central and worst values. We also illustrate the impact of the true central value of $\Delta{m}^2_{31}$ on the sensitivity. In this representation, we show the ratio of the computed sensitivity $\sin^2(2\theta_{13})_{\text{b,w}}$ for the best (resp. worst) case over the sensitivity $\sin^2(2\theta_{13})_{\text{c}}$ for standard central systematic values. The ``Total'' bar shows that in Daya~Bay and RENO the sensitivity can vary from 0.6 to 1.2\,--\,1.3 of the baseline case, for experimental systematics ranging from a ``best'' to a ``worst'' scenario. In the case of Double~Chooz, the impact of systematics is less significant, at the level of 20~\% on both sides. In the Double~Chooz and DB~Mid cases, the sensitivity could be worsened for best fit values of $\Delta{m}^2_{31}$ below $2.5\,10^{-3}~\text{eV}^2$. In DB~Full, the sensitivity is quite stable over the current allowed range for $\Delta{m}^2_{31}$ with only a 5~\% effect on sensitivity.

Double~Chooz is an optimized experiment in the sense of robustness with respect to systematics for a goal sensitivity in the 0.02\,--\,0.03 range. Daya~Bay phase~II is adequate to reach a sensitivity at the level of~0.01. However, a simpler experiment for this class of sensitivity would be a scaled-up variant of Double~Chooz, with a very close near site at 150~m and a 1.5~km baseline for the far site. At the Diablo~Canyon power plant~\cite{RWhitePaper}, where two $3.19~\text{GW}_{\text{th}}$ cores are operational and modest civil engineering works would be required, four 20~t detectors (2~near, 2~far) would give a sensitivity of~0.013 after 3~years of data taking.

\section{Conclusion}
\label{sec:conclusion}

In this work we have presented a detailed comparative analysis of the sensitivities to $\sin^{2}(2\theta_{13})$ of upcoming and proposed reactor experiments.

We have first calculated the sensitivities using all available data published by the respective collaborations for the baseline of both the systematical uncertainties and the experimental setups. Our results are generally in good agreement with the sensitivities quoted by the collaborations: 0.054 for Double~Chooz Phase~I; 0.028 for Double~Chooz Phase~II; 0.041 for Daya Bay ``mid''; 0.0089 for Daya Bay ``full''; and 0.021 for RENO. In the case of Daya Bay, we have additionally evaluated the impact of the proposed filling procedure, with pairs of near-far detectors filled with the same scintillator batch and 4 different scintillator batches. If the hydrogen mass fraction is controlled to 0.5~\% between different batches, the sensitivity worsens slightly to 0.0093. We also examined, still for Day Bay, the until-now implicit assumption that the errors of the relative normalizations between the 8 detectors are fully uncorrelated. In reality this is the most optimistic scenario, since a part of these uncertainties may come from site-dependent systematics. In the most pessimistic scenario (all relative normalizations fully correlated for the same site), the Daya Bay sensitivity would be~0.012.

An important result of this work is that the total thermal power available for an experiment, a figure of merit that has been often used as a strong argument to ``rank'' different projects, has a modest impact on the success of an experiment. Large powers are only available in multi-core sites, which are very difficult to monitor. The associated systematics can be overwhelming with respect to the benefit from the statistics. This is very nicely exemplified by the case of RENO, which would reach the same sensitivity with just 2 of its 6 reactors on, and by Daya Bay ``mid'', which results to be just half way between Double~Chooz Phase~I and Double~Chooz Phase~II. We have illustrated how optimal is the use of the available thermal power in each site through a comparison of the real experiments with ideal ``single core equivalent'' setups: Double Chooz and Daya Bay ``full'' are nearly optimal, Daya Bay ``mid'' and RENO do not make full advantage of their huge power. Daya Bay pays for the complexity of its nuclear power plant by the inevitable construction of two near sites.

We have carried out a detailed and unified $\chi^{2}-$pull analysis for all the experiments, under common assumptions for all systematic uncertainties. This study allowed us to compare all projects on an equal footing and evaluating the impact of each single systematic uncertainty on the sensitivity of each experiment. In this calculation the Double Chooz baseline sensitivity is 0.0235 and a single systematic dominates, which is the error of the relative normalization between near and far detector. This shows that taking into consideration its small mass, Double Chooz is an optimized experiment. With respect to the other projects, the Double Chooz sensitivity has a more pronounced dependence on the best fit $\Delta{m}_{13}^{2}$, with $\ssqtt_{\text{lim}}$ ranging from $0.020$ to $0.031$ for the currently $2\,\sigma-$allowed $\Delta{m}_{13}^{2}$ interval.

Daya Bay ``full'' proves to be robust with respect to the size of the systematical errors and to variations of $\Delta{m}_{13}^{2}$ and is the only partly approved project with an achievable sensitivity potential below 0.01. The Daya Bay sensitivity is dominated by the accuracy of the relative normalization between detectors and the degree of correlation existing on a same site between the latter parameters. With the knowledge of the systematics we have today, the Daya Bay sensitivity would be on average between 0.0094 and 0.0123, depending on the degree of correlation of the systematics between detectors on the same site.

Contrary to Double Chooz and Daya Bay, the sensitivity of RENO is largely degraded by the uncertainties of the reactor powers and fuel composition. This, again, shows that the site is not optimal. Nevertheless, due to the large target mass and optimal baseline, a very competitive sensitivity, around~0.02, is achievable.

This pull analysis also shows that the impact of the backgrounds on the $\chi^{2}$ is minor with respect to other systematics. Backgrounds, at least in our simulation, are therefore not critical in any of the analyzed experiments.

Taken into consideration all the above results, we come to a conclusion about the features of the optimal experiment to approach a \ssqtt sensitivity of~0.01. Where by ``optimal'' we mean a robust sensitivity, the simplest configuration, the minimal amount of civil works and the smallest mass. Such an optimal experiment would have a nuclear power plant layout as simple and powerful as Double Chooz; a favorable topography for sufficiently deep underground laboratories, a very close, single near site, a $\sim1.5\,\mbox{km}$ baseline for the far site and a total target mass of about the half of Daya Bay. Diablo Canyon, California, is a good example of an suitable site for such an experiment.

\section*{Acknowledgement}

We are greatful to A.~\textsc{Milsztajn} for the careful proofreading of this article and his useful comments. We wish also to acknowledge M.~\textsc{Lindner} for fruitful discussions on the comparison of the $\theta_{13}$ experiments. It is also a pleasure to thank M.~\textsc{Cribier} and H.~de~\textsc{Kerret} for continuous helpful exchanges.

\section*{Appendix}

In this section we first describe the way we computed event rates, and then the $\chi^2$ analysis implementation with detailed systematic inclusions.

\subsection*{Event rates}

The visible energy inside the detector has a simple expression as a function of the positron energy or neutrino energy of the inverse $\beta$-decay reaction:
\begin{equation}
\Evis=\Epos+m_e\simeq\Enu-\Enu^{\text{thr}}+2\,m_e\;,
\end{equation}
where $\Enu^{\text{thr}}=M_n-M_p+m_e$ is the \nueb threshold energy of the reaction.
The event rates produced in reactor $R$ and recorded in detector $D$ per visible energy bin $\left[E_i;E_{i+1}\right]$ may be written as
\begin{equation}
 N_{i}^{R,D}(\theta_{13},\Delta{m}^2_{31})=
 \int_0^{+\infty}\dd{L}
 \displaystyle\int_{E_i-m_e}^{E_{i+1}-m_e}\dd{\Epos}
     \int_{\Enu^{\text{thr}}}^{+\infty}\dd{\Enu}\; \frac{\partial N_i^{R,D}}{\partial\Epos}(\Epos,\Enu)
\end{equation}
with
\begin{eqnarray}
 \frac{\partial N_i^{R,D}}{\partial\Epos}(\Epos,\Enu) & = &
 \frac{{\cal N}^{R,D}}{4\pi L_{R,D}^2}\, h(L_{R,D},L)\,\sigma(\Enu)\,\phi^{R,D}(\Enu)\,\epsilon(\Epos)\\
 && \times\, R(\Epos,\Enu)\,P_{ee}(\Enu,L,\theta_{13},\Delta{m}^2_{31})\;,\nonumber
\end{eqnarray}
where $h(L_{R,D},L)$, $\sigma(\Enu)$, $\epsilon(\Epos)$, $\phi^{R,D}(\Enu)$, $R(\Epos,\Enu)$ stand for the finite size effect function the \nueb inverse $\beta$-decay reaction cross section, the $e^+$ detection efficiency, the \nueb flux from reactor $R$ in detector $D$ and the energy response of the detector respectively. The generic normalization factor ${\cal N}^{R,D}$ is the product of the experiment life time by the number of available free protons inside the target volume, the global load factor of reactor $R$ and the dead time of detector $D$. The \nueb flux from reactor $R$ in detector $D$ is described in term of the isotope composition as:
\begin{equation}
 \label{eq:phi}
  \phi^{R,D}(\Enu) = P^R \sum_{\ell}
  \frac{C_{\ell}^R}{E_{\ell}^{\text{fis}}}\phi^{R,D}_{\ell}(\Enu)
\end{equation}
where $\ell=\Uf,~\Uh,~\Pn,~\Po$ labels the most important isotopes contributing to the \nueb flux, $C_{\ell}^R$ is the relative contribution of the isotope $\ell$ to the total reactor power ($C_{\ell}^R=\Nfis\Efis/P^R$, and \Nfis is the number of fissions per second of isotope $\ell$), $E_{\ell}^{\text{fis}}$ is the energy release per fission for isotope $\ell$ and $P^R$ is the thermal power of reactor $R$. In eq.~(\ref{eq:phi}), $\phi_{\ell}^{R,D}(\Enu)$ is the energy differential number of neutrinos emitted per fission by the isotope $\ell$, and we adopt the parameterization for the $\phi_{\ell}^{R,D}(\Enu)$ from ref.~\cite{Vogel}. For the $C_{\ell}^R$ we take a typical isotope composition in a nuclear reactor given in eq.~(\ref{eq:mean_composition}).

We take for $P_{ee}$ the oscillation probability expressed in eq.~(\ref{eq:2nuSP}). We assume in this article a constant efficiency of $\epsilon(\Epos)= 80~\%$ and a Gaussian energy response function
\begin{equation}
R(\Epos,\Enu)=\frac{1}{\sqrt{2\pi}\rho(\Epos)}\exp\left(-\left(\frac{\Epos-\Enu+M_n-M_p}{2\rho(\Epos)}\right)^2\right)\;.
\end{equation}
with an energy resolution of $\displaystyle\rho(\Epos)=8~\%/\sqrt{\Epos[MeV]}$. The finite size effect function $h(L_{R,D},L)$ is assumed to be a Dirac distribution (pointlike sources and detectors) in this paper. The total event rates in i$^{\text{th}}$ bin of detector $D$ is then simply expressed as
\begin{equation}
 N_i^D(\theta_{13},\Delta{m}^2_{31}) = \sum_{R=R_1,\ldots,R_{\Nreac}} N_{i}^{R,D}(\theta_{13},\Delta{m}^2_{31})\;.
\end{equation}

\subsection*{$\chi^2$ analysis and systematics inclusion}

The computed event rates, $N_i^{R,D}$, are then included in a $\chi^2$ \textit{pull-approach} analysis~\cite{Fogli}, where correlations between the systematic uncertainties are properly included:
\begin{equation}
  \label{eq:chi2_2}
  \chi^2 = \min_{\left\lbrace\alpha_k\right\rbrace} \left.
  \sum_{\begin{tabular}{>{$\scriptstyle}c<{$}}\\[-6mm]i=1,\ldots,\Nbins\\[-2mm] D=D_1,\ldots,D_{N_d}\end{tabular}}
  \left(\Delta_i^D-\sum_{k=1}^K \alpha_k S_{i,k}^D\right)^2 + \sum_{k,k'=1}^K\alpha_k W_{k,k'}^{-1} \alpha_{k'}
  \right.\;.
\end{equation}
with,
\begin{equation}
  \Delta_i^D=\left(N_i^{^{\star}D}-N_i^D\right)/U_i^D
  \label{eq:Delta_iD2}
\end{equation}
and $U_i^D$ are given by eq.~(\ref{eq:UiD}). The $\alpha_k$ and $S_{i,k}^D$ coefficients are described in Table~\ref{tab:chi2paramtable}.

\begin{table}[htpb]
\begin{center}
\scalebox{1}{
\begin{tabular}[t]{lcccc}
  \hline
  {\bf Error type} & $\mathbf{k}$ & $\mathbf{S_{i,k}^D\times U_i^D}$ & $\mathbf{\alpha_{i,k}^D}$\\
  \hline\hline
  Absolute normalization & 1 & $\sigma_{\text{abs}}N_i^D$ & $\alpha_{\text{abs}}$\\
  \hline
  \multicolumn{4}{@{\hspace{3mm}}l}{Relative normalization ($\Ns=1$)}\\
  \hspace{.12\textwidth}in $D_1$ & $\Ns+1$ & $\sigma_{\text{rel}}N_i^{D_1}$ & $\alpha_{\text{rel}}^{D_1}$ \\
  \hspace{.15\textwidth}\vdots & \vdots & \vdots & \vdots \\
  \hspace{.12\textwidth}in $D_{\Ndet}$ & $\Ns+\Ndet$ & $\sigma_{\text{rel}}N_i^{D_{\Ndet}}$ & $\alpha_{\text{rel}}^{D_{\Ndet}}$\\
  \hline
  Absolute Energy scale & $\Ndet+2$ & $\sigma_{\text{scl}}N_i^D$ & $\alpha_{\text{scl}}$\\
  \hline
  \multicolumn{4}{@{\hspace{3mm}}l}{Relative Energy scale ($\Ns=\Ndet+2$)}\\
  \hspace{.12\textwidth}in $D_1$ & $\Ns+1$ & $\sigma_{\text{scl}}^{D_1} M_i^{D_1}$ & $\alpha_{\text{scl}}^{D_1}$ \\
  \hspace{.15\textwidth}\vdots & \vdots & \vdots & \vdots \\
  \hspace{.12\textwidth}in $D_{\Ndet}$ & $\Ns+\Ndet$ & $\sigma_{\text{scl}}^{D_{\Ndet}} M_i^{D_{\Ndet}}$ & $\alpha_{\text{scl}}^{D_{\Ndet}}$\\
  \hline
  \multicolumn{4}{@{\hspace{3mm}}l}{Backgrounds ($\Ns=2\Ndet+2$)}\\
  \hspace{.05\textwidth}accidentals in $D_1$& $\Ns+1$ & $\sigma_{B_1}^{D_1}B_{1,i}^{D_1}$ & $\alpha_{B_1}^{D_1}$\\
  \hspace{.15\textwidth}\vdots & \vdots & \vdots & \vdots \\
  \hspace{.05\textwidth}accidentals in $D_{\Ndet}$& $\Ns+\Ndet$ & $\sigma_{B_1}^{D_{\Ndet}}B_{1,i}^{D_{\Ndet}}$ & $\alpha_{B_1}^{D_{\Ndet}}$\\
  \hspace{.05\textwidth}cosmogenics in $D_1$& $\Ns+\Ndet+1$ & $\sigma_{B_2}^{D_1}B_{2,i}^{D_1}$ & $\alpha_{B_2}^{D_1}$\\
  \hspace{.15\textwidth}\vdots & \vdots & \vdots & \vdots \\
  \hspace{.05\textwidth}cosmogenics in $D_{\Ndet}$& $\Ns+2\Ndet$ & $\sigma_{B_2}^{D_{\Ndet}}B_{2,i}^{D_{\Ndet}}$ & $\alpha_{B_2}^{D_{\Ndet}}$\\
  \hspace{.05\textwidth}proton recoils in $D_1$& $\Ns+2\Ndet+1$ & $\sigma_{B_3}^{D_1}B_{3,i}^{D_1}$ & $\alpha_{B_3}^{D_1}$\\
  \hspace{.15\textwidth}\vdots & \vdots & \vdots & \vdots \\
  \hspace{.05\textwidth}proton recoils in $D_{\Ndet}$& $\Ns+3\Ndet$ & $\sigma_{B_3}^{D_{\Ndet}}B_{3,i}^{D_{\Ndet}}$ & $\alpha_{B_3}^{D_{\Ndet}}$\\
  \hline
  \multicolumn{4}{@{\hspace{3mm}}l}{Reactor spectrum shape ($\Ns=5\Ndet+2$)}\\
  \hspace{.12\textwidth}in bin~1& $\Ns+1$ & $\sigma_{\text{shp}}N_1^D$ & $\alpha_{\text{shp},1}$\\
  \hspace{.15\textwidth}\vdots & \vdots & \vdots & \vdots \\
  \hspace{.12\textwidth}in bin~\Nbins& $\Ns+\Nbins$ & $\sigma_{\text{shp}}N_{\Nbins}^D$ & $\alpha_{\text{shp},\Nbins}$\\[2mm]
  \hline
  \multicolumn{4}{@{\hspace{3mm}}l}{Reactor power ($\Ns=5\Ndet+\Nbins+2$)}\\
  \hspace{.12\textwidth}from $R_1$ & $\Ns+1$ & $\sigma_{\text{pwr}}^{R_1}N_i^{R_1,D}$ & $\alpha_{\text{pwr}}^{R_1}$\\
  \hspace{.15\textwidth}\vdots & \vdots & \vdots & \vdots \\
  \hspace{.12\textwidth}from $R_{\Nreac}$ & $\Ns+\Nreac$ & $\sigma_{\text{pwr}}^{R_{\Nreac}}N_i^{R_{\Nreac},D}$ & $\alpha_{\text{pwr}}^{R_{\Nreac}}$\\
  \hline
  \multicolumn{4}{@{\hspace{3mm}}l}{Reactor $R$ composition ($\Ns=5\Ndet+\Nbins+\Nreac+2+4(R-1)$)}\\
  \hspace{.12\textwidth}from \Uf & $\Ns+1$ & $\scmp N_i^{ \Uf,D}$ & $\alpha_{\text{cmp},R}^{\Uf}$\\
  \hspace{.12\textwidth}from \Pn & $\Ns+2$ & $\scmp N_i^{ \Pn,D}$ & $\alpha_{\text{cmp},R}^{\Pn}$\\
  \hspace{.12\textwidth}from \Uh & $\Ns+3$ & $\scmp N_i^{ \Uh,D}$ & $\alpha_{\text{cmp},R}^{\Uh}$\\
  \hspace{.12\textwidth}from \Po & $\Ns+4$ & $\scmp N_i^{ \Po,D}$ & $\alpha_{\text{cmp},R}^{\Po}$\\
  \hline
\end{tabular}
}
\caption{Systematic parameters table. We used the following definitions: \Nbins is the number of bins in the reactor spectra, \Ndet is the number of detectors in the experiment, \Nreac is the number of reactor cores in the NPP, \Ns is short for the number of previously defined systematic parameters. For specific values of \sabs, \srel, \sshp, \sbkg, \spwr, \scmp, we refer to the experiment comparison sections~\ref{sec:comparison}\,--\,\ref{sec:discusion}.}\label{tab:chi2paramtable}
\end{center}
\end{table}
The $S_{i,k}^D$ coefficient represents the shift of the $i^{\text{th}}$ bin of detector $D$ spectrum due to a $1\,\sigma$~variation in the $k^{\text{th}}$ systematic uncertainty parameter $\alpha_k$. Most of the systematics are expressed as function of $N_i^{R,D}$ of $B_i^D$ quantities already described previously. The energy scale systematic coefficients in $S_{i,k}^D$ are defined through $M_i^D$ which follows the relation
\begin{equation}
 M_i^D = \sum_{R=R_1,\ldots,R_{\Nreac}} M_{i}^{R,D}\;,
\end{equation}
where
\begin{equation}
 M_{i}^{R,D}=
 \int_0^{+\infty}\dd{L}
 \displaystyle\int_{E_i-m_e}^{E_{i+1}-m_e}\dd{\Epos}
     \int_{\Enu^{\text{thr}}}^{+\infty}\dd{\Enu}\; \frac{\partial^2 N_i^{R,D}}{\partial\Epos^2}(\Epos,\Enu)\;.
\end{equation}

It is often assumed in the pull-approach that $W_{k,k'}=\delta_{k,k'}$. If we keep this definition of $W_{k,k'}$ then we are faced with the problem that the reactor spectrum shape uncertainties may contribute to the absolute normalization error and the fuel composition uncertainties may contribute to the reactor core power errors. If we want to get rid of these free contributions we could use two methods:
\begin{enumerate}
 \item the first one consists to infer that $\sum_{\ell} \alpha_{\text{cmp},\ell}C_{\ell}^R=0$
 \item the second one is to introduce additional weight terms in the $\chi^2$ definition~(\ref{eq:chi2_2}):
$\sum_{R}\left(\sum_{\ell}\alpha_{\text{cmp},\ell}C_{\ell}^R/\varepsilon_{\text{cmp}}\right)^2$
\end{enumerate}
The first method allows disentangling fuel composition from power uncertainties. However, this assumption is a bit too restrictive since in practice the sum $\sum_{\ell} \alpha_{\text{cmp},\ell}C_{\ell}^RP^R=0$ is only constrained at the  knowledge level of the power of reactor~$R$. The second method has the advantage of allowing an estimate of the level of contribution of this systematic to the power uncertainties. Moreover this simply leads to redefining the $W_{k,k'}$ matrix as
\begin{equation}
 W_{k,k'}^{-1} = \delta_{k,k'}
  + \sum_{\ell,\ell',R} I_{\ell,\ell',k,k'}^{\text{cmp},R}
\frac{C_{\ell}^R C_{\ell'}^R}{\varepsilon_{\text{cmp}}^2}
\;,
\end{equation}
with
\begin{equation}
 I_{\ell,\ell',k,k'}^{\text{cmp},R} =
 \begin{cases}
   \delta_{\ell,k-k_0^R}\delta_{\ell',k'-k_0^R} & \text{if } k,\, k' \in\lbrace k_0^R+1,\ldots,k_0^R+5 \rbrace ,\\
 &k_0^R=5\Ndet+\Nbins+\Nreac+2+4(R-1),\\
   0 & \text{otherwise.}
 \end{cases}
\end{equation}

With these definitions, $\varepsilon_{\text{cmp}}$ determines at which level the fuel composition uncertainties are allowed to contribute to the core power errors. One wants typically that fuel composition uncertainties contribute within the allowed region of power uncertainty. Thus,
\begin{equation}
\varepsilon_{\text{cmp}}=\spwr/P^R\;.
\end{equation}

Regarding the fuel composition uncertainties, \scmp may be assessed roughly at the level of \spwr uncertainties:
\begin{equation}
 \scmp^2=2\,-\,3~\%\;.
\end{equation}

\end{document}